\documentclass[10pt,oneside,onecolumn,aps,pra,preprintnumbers,bibnotes,superscriptaddress,amsmath,amssymb]{revtex4-2}

\usepackage[utf8]{inputenc}
\usepackage{graphicx} % Required for inserting images
\usepackage{subcaption}
\usepackage{epstopdf}

\usepackage{amsmath}
\usepackage{amssymb}
\usepackage{amsfonts}
\usepackage{amsthm}
\usepackage{mathtools}
\usepackage{bm}
\usepackage{braket}
\usepackage{physics}
\usepackage{cancel}

\usepackage[usenames,dvipsnames]{xcolor}
\usepackage{ulem}
\usepackage{hyperref}
\usepackage{lipsum}
\usepackage{comment}
\usepackage{tikz}
\usepackage{tikzscale}

\usetikzlibrary{quantikz2,calc,arrows,chains,matrix,positioning,scopes}

\usepackage{letltxmacro}
\LetLtxMacro{\originaleqref}{\eqref}
\renewcommand{\eqref}{Eq.~\originaleqref}

\begin{document}

\title{Squeezing and adiabaticity breaking in time-dependent quantum harmonic oscillators}

\author{Mattia Orlandini}
\affiliation{Department of Physics and Astronomy, University of Florence, 50019 Sesto Fiorentino, Italy.}

\author{Beatrice Donelli}
\affiliation{Istituto Nazionale di Ottica del Consiglio Nazionale delle Ricerche (CNR-INO), Largo Enrico Fermi 6, I-50125 Firenze, Italy.}
\affiliation{European Laboratory for Non-linear Spectroscopy, Universit\'a di Firenze, I-50019 Sesto Fiorentino, Italy.}

\author{Lorenzo Buffoni}
\affiliation{Department of Physics and Astronomy, University of Florence, 50019 Sesto Fiorentino, Italy.}

\author{Stefano Gherardini}
\affiliation{Istituto Nazionale di Ottica del Consiglio Nazionale delle Ricerche (CNR-INO), Largo Enrico Fermi 6, I-50125 Firenze, Italy.}
\affiliation{European Laboratory for Non-linear Spectroscopy, Universit\'a di Firenze, I-50019 Sesto Fiorentino, Italy.}

\begin{abstract}
The quantum harmonic oscillator with time-dependent frequency is a paradigmatic model of driven quantum dynamics and one of the few nontrivial systems that admits an exact analytical solution. In this review paper, we present a unified treatment of the time-dependent oscillator based on the Lewis-Riesenfeld invariant method, Bogoliubov transformations and the Ermakov-Pinney equation. 
We show how these approaches naturally connect to squeezing for the description of excitations production, and to the breakdown of adiabaticity under generic frequency protocols. Exact results for sudden quenches and smooth ramps are discussed in detail. By explicitly bridging invariant methods and squeezing formalism, this review is meant to provide a comprehensive framework for understanding nonequilibrium dynamics in quadratic potentials, with applications ranging from thermodynamics and condensed matter to quantum control theory.
\end{abstract}

\maketitle

%%%%%%%%%%%%%%%%%%%%%%%%%%%%%%%%%%%%%%%%
\section{Introduction}

The quantum harmonic oscillator (QHO) with a time-dependent frequency is one of the most paradigmatic models in quantum physics. It provides a rare example of a nonstationary quantum system whose dynamics can still be treated exactly. It indeed captures the universal physical behavior of a broad variety of driven many-body and field-theoretic settings, ranging from shortcuts to adiabaticity~\cite{Chen_STA,GuryOdelin2019} and quantum thermodynamics~\cite{Deffner_2008,AbahDeffner2010,DelCampo2014,Abah2012SingleIonHeatEngine,Abah2018STAquantumEngine,GomezRuizQST2025}, to condensed matter and phase transitions~\cite{PueblaQuantumRabiModel,PueblaPlenioSuperradiantQPT,Defenu_2018,Defenu_2021QAC}. Concerning the applications, we can mention optomechanical systems~\cite{Barzanjeh2022,Zhang2014}, models (as the Bose-Hubbard one) to which the Holstein-Primakoff approximation can be applied~\cite{LenaPRA2016,Defenu_2018,Buffoni_Gherardini,gyamfi2019introductionholsteinprimakofftransformationapplications,Dusuel2005}, QED vacuum pair production in a time dependent electromagnetic field (Schwinger effect)~\cite{Scwinger1951,Greiner1985,Popov1969} or particle productions in expanding universe models~\cite{Parker1,Parker2}.

The issue of determining the solution of a time-dependent QHO has been addressed more than once, using different approaches. Among the most widely used, it is worth considering the Ermakov-Pinney equation~\cite{ermakov1880,pinney1950}, the Lewis-Riesenfeld dynamical invariant method~\cite{L-R1969,Lewis1967}, and even the use of Bogoliubov transformations~\cite{ring2004nuclear,blaizot1986quantum}. Each of these approaches leverages a different aspect of the dynamics' solution: the Ermakov equation encodes the non-trivial time evolution of the system's wave-packet width; invariants provide an exact basis over which the solution of the time-dependent Schr\"{o}dinger equation can be expressed, while Bogoliubov coefficients showcase the analytical expression for how the annihilation and creation operators decomposing the driven Hamiltonian of the system are mixed.

In this review, devoted to undergraduate and graduate students, rather than focusing on invariant methods that have been recently reviewed in Ref.~\cite{CoelhoReview2025}, we pedagogically discuss how and to which extent a generic time-dependent protocol breaks the \textit{quantum adiabatic theorem}~\cite{messiah2014quantum,kato1950adiabatic}, as an effect of excitations' creation. When the system's Hamiltonian is varied slowly, the state of the system is one of the eigenstates of the Hamiltonian at any time. On the contrary, if the driving protocols is performed faster, the state of the system is in a superposition of the instantaneous eigenstates of the Hamiltonian with the coefficients of the superposition depending on the specific initial state. The latter condition is denoted as non-adiabatic, and the projection of the system's state over the Hamiltonian basis gives rise to the so-called non-adiabatic excitations. Interestingly, for the QHO, the dynamical transformation that changes the state of the system under a generic time-dependent protocol is exactly a \textit{squeezing} operator~\cite{loudon2000quantum,barnett1997methods,Husimi}. The latter is built over the time-dependent annihilation and creation operators that are associated with the system's Hamiltonian at the current time of the transformation. Such an association is allowed by Bogoliubov transformations.

In the present review, we present a {\it unified formulation} that keeps explicit the most relevant mutual relations between adiabaticity breaking and squeezing generation. In this way, we aim to clarify how the same solution for the dynamics of a QHO under a generic driving protocol can be read simultaneously as an invariant evolution, or as the result of a Bogoliubov transformation, or even as a squeezing process generated by the protocol. For the purposes explained above, we express the operator of squeezing in the instantaneous Hamiltonian basis, as a function of the harmonic frequency $\omega(t)$, at any time $t$. Then, the amount of squeezing is evaluated on a time-dependent basis. This choice allows one to identify non-adiabatic excitations, due to the driving protocol, directly with the departure of the driven system from the instantaneous ground-state manifold, making clear the connection between adiabaticity breaking and squeezing.

The review paper is organized as follows. First, in Secs.~\ref{sec:LR_method}-\ref{sec:QHO} we review the Lewis-Riesenfeld invariant method in its general form and then we specialize it to the one-dimensional QHO with time-dependent frequency. Then, in Secs.~\ref{squeezinginterpretationistantaneus}, \ref{Section:Exitationampl}, \ref{sec:symmetries}, \ref{section: AdiabaticFactor}, we connect this framework to Bogoliubov transformations and squeezing operators, by discussing the role of the instantaneous Hamiltonian basis, and the breaking of the quantum adiabatic theorem. Moreover, in Sec.~\ref{sec:solution_Ermakov} we report the solution of the Ermakov equation, while in Sec.~\ref{sec:protocols} we give the solution of two relevant families of driving protocols, i.e., sudden quenches and smooth frequency variations through a linear ramp. Finally, technical derivations are reported in Appendices. 

%%%%%%%%%%%%%%%%%%%%%%%%%%%%%%%%%%%%%%%%
\section{Lewis and Riesenfeld invariant method}
\label{sec:LR_method}

Let us consider a system whose Hamiltonian operator $\hat{H}(t)$ depends explicitly on time. This means that the system is not isolated and its energy changes over time, for example, by tuning a control parameter under a specific protocol.

The Lewis-Riesenfeld dynamical invariant method~\cite{L-R1969,Lewis1967} gives a procedure to construct exact solutions of the Schr\"{o}dinger equation for driven quantum systems, whose evolution is governed by the differential equation
\begin{equation}\label{SE}
i \hbar\frac{\partial }{\partial t} \ket{\psi\left( t \right)}=\hat{H}\left( t \right)\ket{\psi\left( t \right)}.
\end{equation}
In the following, we will outline this method in the general case, explaining its fundamental ideas and steps. Then, we will apply it to a one-dimensional QHO with a time-dependent frequency. 

\subsection{General method}

Following the original work in Ref.~\cite{L-R1969} and the recent review \cite{CoelhoReview2025}, we assume the existence of a Hermitian operator $\hat{I}(t)$ that depends explicitly on time and is an invariant, i.e.,
\begin{equation}\label{DI}
    \frac{1}{i\hbar}\left[ \hat{I}(t),\hat{H}(t)  \right]+\frac{\partial \hat{I}(t)}{\partial t}=0.
\end{equation}
\eqref{DI} entails that the invariant $\hat{I}(t)$ is a constant of motion in the Heisenberg picture (denoted with the superscript $H$ from here on), since it satisfies the equality $\frac{d\hat{I}^{H}(t)}{dt}=0$.
In order to fix the notation, we briefly recall the connection between the Heisenberg and Schr\"{o}dinger pictures~\cite{konishi,Sakurai,messiah2014quantum}. Let $\hat{U}(t,0)$ denote the unitary operator for the time evolution of the system from $t=0$ to the generic time $t$, associated with the Hamiltonian $\hat{H}(t)$. If $\ket{\psi(t)}$ is the state of the system at the generic time $t$, the Heisenberg picture is defined by the following unitary transformation:
\begin{equation}
    \ket{\psi(t)} \longrightarrow \ket{\psi^{H}}=\hat{U}^{\dagger}(t,0)\ket{\psi(t)}.
\end{equation}
This transformation transfers the dynamics from the state to the observables.
In particular, for a generic Hermitian operator $\hat{A}(t)$, we have that~\cite{konishi,Sakurai,messiah2014quantum}
\begin{equation}
    \hat{A}^{H}\left( t \right)=\hat{U}^{\dagger}(t,0)\hat{A}\left( t \right)\hat{U}\left( t,0 \right),
\end{equation}
which is such that
\begin{equation}
   \frac{d \hat{A}^{H}(t)}{dt}=\frac{1}{i\hbar}\left[ \hat{A}^{H}(t),\hat{H}^{H}(t) \right]+\frac{\partial \hat{A}^{H}(t)}{\partial t}.
\end{equation}
Applying the left-hand side of \eqref{DI} to the state $\ket{\psi(t)}$ and using \eqref{SE}, we obtain:
\begin{equation}\label{Eq2_on_psi}
        \frac{\partial \hat{I}(t)}{\partial t}\ket{\psi(t)}=\frac{1}{i\hbar}\hat{H}(t)\hat{I}(t)  \ket{\psi(t)} - \hat{I}(t)\frac{\partial }{\partial t}\ket{\psi(t)}.
\end{equation}
\eqref{Eq2_on_psi} can be re-written as
\begin{equation}
    i\hbar\frac{\partial }{\partial t}\Big( \hat{I}(t)\ket{\psi(t)} \Big) = \hat{H}(t)\Big( \hat{I}(t)\ket{\psi\left( t \right)}\Big),
\end{equation}
which implies that the action of the invariant $\hat{I}(t)$ on a solution of the Schr\"{o}dinger equation returns another state that satisfies \eqref{SE}. This result is valid for a generic invariant.

In the following, we will assume that $\hat{I}(t)$ does not contain time derivatives and we will show that there exist simple and explicit rules to determine the phases of the invariant's eigenstates such that \eqref{SE} is satisfied. We denote the eigenvalues of $\hat{I}(t)$ as $\lambda$ and the relative eigenkets as $\ket{\lambda,\eta;t}$, where $\eta$ is the index denoting degeneracy. We assume that the eigenstates of $\hat{I}(t)$ form an orthonormal basis of the system's Hilbert space, and in first approximation that the spectrum is discrete. Therefore, 
\begin{equation}\label{Ieigen}
\hat{I}(t)\ket{\lambda,\eta;t}=\lambda\ket{\lambda,\eta;t},
\end{equation}
such that $\sum_{\lambda,\eta}^{}\ket{\lambda,\eta;t}\bra{\lambda,\eta;t}=\hat{\mathbb{I}}$ and $\langle\lambda,\eta;t|\lambda',\eta';t\rangle=\delta_{\lambda,\lambda'}\delta_{\eta,\eta'}$ for any time $t$.

An important consequence of the invariance property \eqref{DI} is that {\it the eigenvalues $\lambda$ are time-independent}.
In order to show this, we apply \eqref{DI} to $\ket{\lambda,\eta;t}$ and we take the scalar product with the state $|\lambda',\eta';t\rangle$. Then, using \eqref{Ieigen}, we get:
\begin{equation}\label{matrixderI1}
    \langle\lambda',\eta';t|\frac{\partial \hat{I}(t)}{\partial t}\ket{\lambda,\eta;t}=\frac{(\lambda-\lambda')}{i\hbar}\langle\lambda',\eta';t|\hat{H}(t)\ket{\lambda,\eta; t}.
\end{equation}
Now we take the time derivative of \eqref{Ieigen} considering a possible time-dependence of the eigenvalues $\lambda$:
\begin{equation}\label{DerEigenEq}
    \frac{\partial \hat{I}(t) }{\partial t}\ket{\lambda,\eta;t}+\hat{I}(t)\frac{\partial }{\partial t}\ket{\lambda,\eta;t}=\frac{\partial \lambda }{\partial t}\ket{\lambda,\eta;t}+\lambda\frac{\partial  }{\partial t}\ket{\lambda,\eta;t}.
\end{equation}
Multiplying both sides of \eqref{DerEigenEq} by the bra $\langle\lambda',\eta';t|$, we have that
\begin{equation}\label{matrixderI2}
    \langle\lambda',\eta';t|\frac{\partial \hat{I}(t)}{\partial t}\ket{\lambda,\eta;t}=(\lambda-\lambda') \langle\lambda',\eta';t|\frac{\partial }{\partial t} \ket{\lambda,\eta;t} +\delta_{\lambda,\lambda'}\delta_{\eta,\eta'}\frac{\partial \lambda }{\partial t}.
\end{equation}
Thus, equating the right-hand sides of \eqref{matrixderI1} and \eqref{matrixderI2}, we obtain:
\begin{equation}\label{derivs_equality}   \delta_{\lambda,\lambda'}\delta_{\eta,\eta'}\frac{\partial \lambda }{\partial t} + (\lambda-\lambda') \langle\lambda',\eta';t|\frac{\partial }{\partial t}\ket{\lambda,\eta;t} = \frac{(\lambda-\lambda')}{i\hbar}\langle\lambda',\eta';t|\hat{H}(t)\ket{\lambda,\eta;t}.
\end{equation}
\eqref{derivs_equality} must be true for every $\lambda,\lambda'$ and $\eta,\eta'$; thus, we have to require that $\delta_{\lambda,\lambda'},\delta_{\eta,\eta'}\neq 0$ choosing $\lambda=\lambda'$ and $\eta=\eta'$, such that
\begin{equation}\label{TimeIndEigen}
\frac{\partial\lambda }{\partial t }=0 \,.
\end{equation}

At this point, let us determine the relationship between the solution of the Schr\"{o}dinger equation and the eigenstates of $\hat{I}(t)$. In order to do so, we use \eqref{TimeIndEigen} to rewrite \eqref{DerEigenEq} as
\begin{equation}
      \frac{\partial \hat{I}(t)}{\partial t}\ket{\lambda,\eta;t} = \left( \lambda - \hat{I}(t) \right)\frac{\partial  }{\partial t}\ket{\lambda,\eta; t}.
\end{equation}
Taking the scalar product with $|\lambda',\eta';t\rangle$ and using \eqref{matrixderI1} with $\langle\lambda',\eta';t|\hat{I}(t)=\lambda'\langle\lambda',\eta';t|$, we determine:
\begin{equation}\label{HIrelation}
    (\lambda-\lambda')\langle\lambda',\eta';t|\Big( \hat{H}(t)-i\hbar\frac{\partial }{\partial t} \Big)\ket{\lambda,\eta; t}=0.
\end{equation}
For $\lambda\neq\lambda'$,
\begin{equation}\label{HIrelation1}
   i\hbar\langle\lambda',\eta';t|\frac{\partial }{\partial t}\ket{\lambda,\eta,t}=\langle\lambda',\eta';t|\hat{H}(t)\ket{\lambda,\eta;t}.
\end{equation}

It is worth noting that if \eqref{HIrelation1} was valid also for $\lambda=\lambda'$, then we would deduce that $\ket{\lambda,\eta;t}$ is a solution of \eqref{SE}, but it is not necessarily valid. Indeed, as Lewis and Riesenfeld observed~\cite{L-R1969,Lewis1967}, one is free to multiply $\ket{\lambda,\eta;t}$ by an arbitrary time-dependent phase factor. In this way, we can define a new set of eigenstates of the invariant through a time-dependent gauge transformation:
\begin{equation}\label{neweigen}    \ket{\lambda,\eta;t}_{\alpha}=e^{i\alpha_{\lambda,\eta}(t)}\ket{\lambda,\eta;t},
\end{equation}
where $\alpha_{\lambda,\eta}(t)$ are real time-dependent functions.
Since $\hat{I}(t)$ is assumed to not contain any time-derivative, the states $\ket{\lambda,\eta;t}_{\alpha}$ are still orthonormal eigenkets of $\hat{I}(t)$, corresponding to the same eigenvalues associated to $\ket{\lambda,\eta;t}$:
\begin{equation}\label{Ieigenalpha}    
\hat{I}(t)\ket{\lambda,\eta;t}_{\alpha}=\lambda\ket{\lambda,\eta;t}_{\alpha}.
\end{equation}
Now we can impose that also the eigenkets $\ket{\lambda,\eta;t}_{\alpha}$ satisfy \eqref{HIrelation1} by properly choosing the phases $\alpha_{\lambda,\eta}(t)$. This is equivalent to require the validity of the following equation:
\begin{equation}\label{HIrelation1_alpha}
{}_{\alpha}\!\bra{\lambda',\eta';t}i\hbar\frac{\partial }{\partial t}\ket{\lambda,\eta;t}_{\alpha} = {}_{\alpha}\!\bra{\lambda',\eta';t}\hat{H}(t)\ket{\lambda,\eta;t}_{\alpha}
\end{equation}
for generic $\lambda, \lambda'$. Using \eqref{neweigen} and making explicit the time-dependence of $\ket{\lambda,\eta;t}$, the left-hand side of \eqref{HIrelation1_alpha} is
\begin{equation}\label{HIrelation1_alpha_2}
{}_{\alpha}\!\langle\lambda',\eta';t|i\hbar\frac{\partial }{\partial t}\ket{\lambda,\eta;t}_{\alpha} = e^{ i\left(  \alpha_{\lambda,\eta}(t)-\alpha_{\lambda',\eta'}(t)\right) }\left( -\hbar \delta_{\lambda,\lambda'}\delta_{\eta,\eta'}\frac{d\alpha_{\lambda,\eta}(t) }{dt} + \bra{\lambda',\eta';t}i\hbar\frac{\partial }{\partial t}\ket{\lambda,\eta;t}\right).
\end{equation}
On the other hand, using \eqref{Ieigenalpha}, the right-hand side of \eqref{HIrelation1_alpha} simplifies to
\begin{equation}
    {}_{\alpha}\!\bra{\lambda',\eta';t}\hat{H}(t)\ket{\lambda,\eta;t}_{\alpha} = e^{ i\left(  \alpha_{\lambda,\eta}(t)-\alpha_{\lambda',\eta'}(t)\right) } \bra{\lambda',\eta';t}\hat{H}(t)\ket{\lambda,\eta;t}{}.
\end{equation}
Therefore, \eqref{HIrelation1_alpha} can be re-written as 
\begin{equation}\label{Fundamentaleq}    \hbar\delta_{\lambda,\lambda'}\delta_{\eta,\eta'}\frac{d \alpha_{\lambda,\eta}(t)}{dt} = \bra{\lambda',\eta';t}\left( i\hbar \frac{\partial }{\partial t}-\hat{H}(t)\right)\ket{\lambda,\eta;t}.
\end{equation}
For $\lambda\neq\lambda'$, \eqref{Fundamentaleq} holds thanks to the validity of \eqref{HIrelation1}. Instead, in order to satisfy \eqref{Fundamentaleq} when $\lambda=\lambda'$ and $\eta\neq\eta{'}$, the states $\ket{\lambda,\eta;t}$ must be chosen such that the right-hand side of \eqref{Fundamentaleq} vanishes, or in other words the operator $i\hbar\frac{\partial }{\partial t}-\hat{H}(t)$ must be diagonalizable in the eigenspace associated with the eigenket $\ket{\lambda,\eta;t}$
and eigenvalue $\lambda$. This property can be fulfilled since the operator $i\hbar\frac{\partial }{\partial t}-\hat{H}(t)$ is Hermitian. Hence, for $\lambda=\lambda'$, we have that
\begin{equation}\label{PhasesEquation}
    \hbar\frac{d \alpha_{\lambda,\eta}(t)}{dt}=\bra{\lambda,\eta;t}\left( i\hbar \frac{\partial }{\partial t}-\hat{H}(t)\right)\ket{\lambda,\eta;t}.
\end{equation}

By construction, the new set of eigenkets $\ket{\lambda,\eta;t}_{\alpha}$ satisfy the Schr\"{o}dinger equation \eqref{SE}, and any solution of \eqref{SE} can be obtained as a linear combination of these states. 
Accordingly, having denoted with $\ket{\psi(t)}$ the solution of \eqref{SE}, it can be expressed in the form
\begin{equation}\label{General solution}
\ket{\psi(t)}=\sum_{\lambda,\eta}^{}c_{\lambda,\eta} \, e^{i\alpha_{\lambda,\eta}(t)}\ket{\lambda,\eta;t},    
\end{equation}
where the coefficients $c_{\lambda,\eta}$ depends on the initial state $\ket{\psi(0)}$ as $c_{\lambda,\eta}=\bra{\lambda,\eta;0}\ket{\psi(0)}$. 
Notice that \eqref{General solution} can be formally rewritten as
\begin{equation}    
\ket{\psi(t)} = \left( \sum_{\lambda,\eta}^{}e^{i\alpha_{\lambda,\eta}(t)}\ket{\lambda,\eta;t}\!\bra{\lambda,\eta;0} \right) \ket{\psi(0)},
\end{equation}
from which we can deduce a formal expression for the time-evolution unitary operator that is associated with the Hamiltonian $\hat{H}(t)$, i.e.,
\begin{equation}\label{eq:InvariantEvolutionOperator}
    \hat{U}(t,0)=\sum_{\lambda,\eta}^{}e^{i\alpha_{\lambda,\eta}(t)}\ket{\lambda,\eta;t}\!\bra{\lambda,\eta;0}.
\end{equation}

In conclusion, we can summarize the Lewis-Riesenfeld (LR) method through the following steps \cite{CoelhoReview2025}:
\begin{itemize}
    \item 
    we construct an invariant operator $\hat{I}(t)$ that obeys \eqref{DI};
    \item 
    we determine the eigenvalues and eigenstates of $\hat{I}(t)$ by solving the eigenvalue problem
    \begin{equation}
    \hat{I}(t)\ket{\lambda,\eta;t}=\lambda\ket{\lambda, \eta;t};
    \end{equation}
    \item 
    after have obtained the eigenkets $\ket{\lambda,\eta;t}$, we compute the phase functions $\alpha_{\eta\lambda}(t)$ solving the differential equation
    \begin{equation}
        \hbar\frac{
        d \alpha_{\lambda,\eta}(t)}{dt}=\bra{\lambda,\eta;t}\left( i\hbar \frac{\partial }{\partial t}-\hat{H}(t)\right)\ket{\lambda,\eta;t};
    \end{equation}
    \item 
    given the initial state $\ket{\psi(0)}$, a general solution $\ket{\psi(t)}$ of the Schr\"{o}dinger equation is obtained from \eqref{General solution}.
\end{itemize}

%%%%%%%%%%%%%%%%%%%%%%%%%%%%%%%%%%%%%%%%
\section{Analytical dynamics of a time-dependent quantum harmonic oscillator}
\label{sec:QHO}

In this section, we show how the steps of the LR method described above can be analytically carried out for a QHO with time-dependent frequency $\omega(t)$ and constant mass $M$. The Hamiltonian operator of this system is
\begin{equation}\label{OscillatorHam}
    \hat{H}(t)=\frac{\hat{p}^2}{2M}+\frac{1}{2}M\omega^2(t)\hat{q}^2 .
\end{equation}
The operators $\hat{q}$ and $\hat{p}$ satisfy the canonical commutation relation $[\hat{q},\hat{p}]=i\hbar$. Thus, if we introduce the operators
\begin{equation}
    \hat{K}_0=\frac{\left\{ \hat{q},\hat{p} \right\}}{2},\qquad\hat{K}_-=\frac{\hat{p}^2}{2},\qquad\hat{K}_+=\frac{\hat{q}^2}{2},
\end{equation}
where $\left\{ \hat{q}, \hat{p}\right\} = \hat{q}\hat{p} + \hat{p}\hat{q}$ denotes the anti-commutator of $\hat{q}$ and $\hat{p}$, then they satisfy the commutation rules
\begin{equation}\label{SU(1,1) algebra}
    \left[ \frac{i}{2\hbar}\hat{K}_0,\hat{K}_\pm  \right]=\pm \hat{K}_{\pm} \quad \text{and} \quad \left[  \hat{K}_+,\hat{K}_-\right]=2i\hbar\left( \frac{\hat{K}_0}{2} \right).
\end{equation}
The Hamiltonian operator (\ref{OscillatorHam}) can be rewritten as
\begin{equation}
    \hat{H}(t)=\frac{1}{M}\hat{K}_-\ +\frac{\omega^2(t)}{M}\hat{K}_{+}.
\end{equation}

We are going to search for an Hermitian invariant of the following form:
\begin{equation}\label{InvariantAnsatz}
  \hat{I}(t)=\alpha(t)\hat{K}_++\beta(t)\hat{K}_- + \gamma(t)\hat{K}_{0},
\end{equation}
where $\alpha(t), \,\beta(t), \, \gamma(t)$ are real functions of time.
Substituting the ansatz (\ref{InvariantAnsatz}) for $\hat{I}(t)$ in the invariance condition of \eqref{DI} and using the commutation relations (\ref{SU(1,1) algebra}), we get a system of coupled ordinary differential equations that involve $\alpha(t)$, $\beta(t)$ and $\gamma(t)$: 
\begin{equation}\label{System}
\begin{cases}
\displaystyle{\dot\alpha(t)=2M\omega^2(t)\gamma(t)}\\ \\
\displaystyle{\dot\beta(t)=-\frac{2\gamma(t)}{M}} \\ \\
\displaystyle{\dot\gamma(t)=-\frac{\alpha(t)}{M}+M\omega^2(t)\beta(t)}.
\end{cases}
\end{equation}
In order to simplify this system of equations, it is convenient to introduce the auxiliary function $\sigma(t)$ defined by
\begin{equation}
    \beta(t)=\sigma^2(t).
\end{equation}
From the second equation of \eqref{System}, we can write:
\begin{equation}
\gamma(t) = -M\sigma(t)\dot\sigma(t).
\end{equation}
Substituting the latter in the third equation of (\ref{System}), we get an expression of $\alpha(t)$ in terms of $\sigma(t)$ and its derivatives:
\begin{equation} \label{eq:alpha_t}   
\alpha(t)=M^2\dot\sigma^2(t)+M^2\sigma(t)\Big( \ddot\sigma(t)+\omega^2(t)\sigma(t) \Big).
\end{equation}
Replacing \eqref{eq:alpha_t} in the first equation of (\ref{System}), we obtain a constraint on $\sigma(t)$, expressed by the following differential equation:
\begin{equation}    3\dot\sigma(t)\Biggl(\ddot\sigma(t)+\omega^2(t)\sigma(t)\Biggl)+\sigma(t)\frac{d}{dt}\Biggl(\ddot\sigma(t)+\omega^2(t)\sigma(t)\Biggl)=0.
\end{equation}
Setting $x(t) = \ddot\sigma(t)+\omega^2(t)\sigma(t)$, we 
end-up with the differential equation $3\dot\sigma(t) x(t) + \sigma(t)\dot{x}(t)=0$, which we can integrate by separation of variables, obtaining $\ln{|x(t)|}=\ln{|\sigma(t)^{-3}|}+c_1$ that is $|x(t)|=c_2/|\sigma(t)^{3}|$, with $c_1,c_2$ real constants of integration. Overall, we get:
\begin{equation}\label{EPequation1}
\ddot\sigma(t)+\omega^2(t)\sigma(t)=\frac{c}{\sigma^{3}(t)},
\end{equation}
where also $c$ is a real constant of integration. Solving the differential equation (\ref{EPequation1}) leads us to determine the time-behavior of $\alpha(t)$, $\beta(t)$, $\gamma(t)$ that we have expressed in terms of $\sigma(t)$ and its derivatives:
\begin{equation}\label{sigmaeq}
\begin{cases}
\displaystyle{\alpha(t)=M^{2}\dot\sigma^{2}(t)+\frac{c M^{2}}{\sigma^{2}(t)}}\\
\displaystyle{\beta(t)=\sigma^{2}(t)}\\
\displaystyle{\gamma(t)=-M\sigma(t)\dot\sigma(t)}.
\end{cases}
\end{equation}
Using \eqref{sigmaeq}, the invariant of \eqref{InvariantAnsatz} becomes
\begin{equation}\label{Invariant_c}
    \hat{I}(t)=\frac{1}{2}\left( \frac{c\,M^{2}}{\sigma^{2}(t)}\hat{q}^{2}+\Big( \sigma(t) \hat{p} - M\dot\sigma(t) \hat{q} \Big)^{2} \right)
\end{equation}
with $\sigma$ satisfying \eqref{EPequation1}.
The presence of the arbitrary constant of integration $c$ in Eqs.~(\ref{EPequation1}) and (\ref{Invariant_c}) allows for scale transformations of the auxiliary variable $\sigma(t)$, making it possible to derive different invariants that are equivalent up to constant multiplication.
This justifies the different types of invariants present in the current literature~\cite{Lewis1967,L-R1969,Ji_1995_01,Ji_1995_02,Pedrosa1997,FernandezGuasti2003,Chen_STA,DelCampo2014,DittrichReuter2020}.
In the following, we are going to use the invariant
\begin{equation}\label{LR_Invariant}
\hat{I}(t) = \frac{1}{2}\left( \frac{\hat{q}^{2}}{\sigma^{2}(t)}+\Big( \sigma(t) \hat{p} - M\dot\sigma(t) \hat{q}\Big)^{2} \right),
\end{equation}
where we have set $c=1/M^{2}$ and $\sigma(t)$ satisfies the differential equation
\begin{equation}\label{EPequation2}
\ddot\sigma(t)+\omega^2(t)\sigma(t)=\frac{1}{M^{^{2}}\sigma^{3}(t)}.
\end{equation}
Notice that \eqref{LR_Invariant} has the dimension of an action (i.e., Joule x seconds).

\subsection{Spectrum and eigenstates of the invariant}

Having built an Hermitian invariant for the time-dependent QHO, we now determine its spectrum and eigenstates. To do this, we introduce the following time-dependent annihilation and creation operators:
\begin{eqnarray} \label{cd_definition}
\begin{cases}
\displaystyle{\hat{a}(t) = \frac{1}{\sqrt{2\hbar}}\left( \frac{\hat{q}}{\sigma(t)} + i\Big( \sigma(t)\hat{p} - M\dot{\sigma}(t)\hat{q} \Big)\right)} \\
\displaystyle{\hat{a}^{\dagger}(t) = \frac{1}{\sqrt{2\hbar}}\left( \frac{\hat{q}}{\sigma(t)} - i\Big( \sigma(t)\hat{p} - M\dot{\sigma}(t)\hat{q}\Big) \right)},
\end{cases}
\end{eqnarray}
such that the invariant (\ref{LR_Invariant}) in terms of $\hat{a}(t)$ and $\hat{a}^{\dagger}(t)$ assumes the form
\begin{equation}\label{Invariant_cd}
\hat{I}(t)=\hbar\left( \hat{a}^{\dagger}(t)\hat{a}(t)+\frac{1}{2}\right). 
\end{equation}
Notice that, using the canonical commutation relations of $\hat{q}$ and $\hat{p}$, it holds that $\left[ \hat{a}(t), \hat{a}^{\dagger}(t)\right] = \hat{\mathbb{I}}$.

Resorting to \eqref{Invariant_cd}, we have a formal analogy with the QHO with constant (time-independent) frequency. We can thus determine the time-dependent eigenstates and eigenvalues of the invariant $\hat{I}(t)$ following the standard operator technique~\cite{konishi,Sakurai,messiah2014quantum}.
The eigenstates $\ket{n;t}_{I}$ of the invariant in \eqref{LR_Invariant} are the same of the number operator $\hat{a}^{\dagger}(t)\hat{a}(t)$, i.e.,
\begin{equation}
\hat{a}^{\dagger}(t)\hat{a}(t)\ket{n;t}_{I}=n\ket{n;t}_{I} \quad \text{with} \quad n=0,1,...
\end{equation}
Hence, from \eqref{Invariant_cd}, we have:
\begin{equation}
\hat{I}(t)\ket{n;t}_{I} = \hbar\left( n+\frac{1}{2} \right)\ket{n;t}_{I}\qquad n=0,1,...,
\end{equation}
where the action of $\hat{a}(t)$ and $\hat{a}^{\dagger}(t)$ on the eigenkets of $\hat{I}(t)$ is given by:
\begin{eqnarray}
\hat{a}(t)\ket{n;t}_{I} &=&\sqrt{n}\ket{n-1;t}_{I}\label{action_a} \\
\hat{a}^{\dagger}(t)\ket{n;t}_{I} &=&\sqrt{n+1}\ket{n+1;t}_{I}.\label{action_a_dag}
\end{eqnarray}
As a result, the eigenstate $\ket{n;t}_{I}$ of $\hat{I}(t)$ with eigenvalue $\lambda_{n}=\hbar\left(n+\frac{1}{2}\right)$ is:
\begin{equation}\label{InvariantFockeigenstates}
\ket{n;t}_{I}=\frac{\Big( \hat{a}^{\dagger}(t)\Big)^n }{\sqrt{n!}}\ket{0;t}_{I}.
\end{equation}

\subsection{Phase terms associated with the eigenstates of the invariant}\label{PhasetermsSection}

The arbitrary non-adiabatic~\cite{Ji_1995_01,Ji_1995_02} Berry phase terms solving \eqref{PhasesEquation} for a time-dependent QHO can be written as 
\begin{equation}\label{Phase_ode}
\hbar\frac{d\alpha_{n}(t)}{dt}= {}_{I}\!\bra{n;t}\left( i\hbar\frac{\partial}{\partial t}-\hat{H}(t)\right)\ket{n;t}_{I}.
\end{equation}
Thus, we have to evaluate the diagonal matrix elements of $\hat{H}(t)$ and $\frac{\partial}{\partial t}$, i.e., ${}_{I}\!\bra{n;t}\hat{H}(t)\ket{n;t}_I$ and ${}_{I}\!\bra{n;t}\frac{\partial}{\partial t}\ket{n;t}_I$. In order to compute the former, we use \eqref{cd_definition} to express the Hamiltonian \eqref{OscillatorHam} in terms of the annihilation and creation operators (see App.~\ref{app:definitions_L}):
\begin{equation}\label{Hamiltonian_InvariantOperatorsExpression}
\hat{H}(t) = \frac{\hbar}{2}\Big(
\hat{L}_0(t) + \hat{L}_{-}(t) + \hat{L}_{+}(t)\Big),
\end{equation}
where the operators $\hat{L}_0$, $\hat{L}_{-}$, $\hat{L}_{+}$ are defined as 
\begin{equation}\label{definitions_L}
\begin{cases}
\displaystyle{ \hat{L}_0(t) = \frac{1}{2}\left( \frac{1}{M\sigma^2(t)}+M\omega^2(t)\sigma^2(t)+M\dot{\sigma}^2(t)\right)\left\lbrace \hat{a}(t), \hat{a}^{\dagger}(t)\right\rbrace}\\
\displaystyle{ \hat{L}_{-}(t)=\frac{1}{2}\left( M\left(\dot{\sigma}^2(t) -\sigma(t)\ddot{\sigma}(t)\right)-2i\frac{\dot{\sigma}(t)}{\sigma(t)}\right)\,\hat{a}^2(t)}\\
\displaystyle{ \hat{L}_{+}(t)=\frac{1}{2}\left( M\left(\dot{\sigma}^2(t) -\sigma(t)\ddot{\sigma}(t)\right)+2i\frac{\dot{\sigma}(t)}{\sigma(t)}\right)\hat{a}^{\dagger2}(t).}
\end{cases} 
\end{equation}
From the expressions in \eqref{definitions_L}, the only operator in $\hat{H}(t)$ that has non-zero diagonal matrix elements in the basis spanned by $\ket{n;t}_I$ is $\hat{L}_0(t)$. In particular,
\begin{equation}\label{averageH}
{}_{I}\!\bra{n;t}\hat{H}(t)\ket{n;t}_I = \frac{\hbar}{2}\left(n+ \frac{1}{2}\right)\left( \frac{1}{M\sigma^2(t)}+M\omega^2(t)\sigma^2(t)+M\dot{\sigma}^2(t) \right).
\end{equation}
Then, concerning ${}_{I}\!\bra{n;t}\frac{\partial}{\partial t}\ket{n;t}_I$, from \eqref{action_a_dag} we have that $\hat{a}^{\dagger}(t)\ket{n-1;t}_{I}=\sqrt{n}\ket{n;t}_{I}$, such that 
\begin{equation}\label{diag_element_time-derivative}
{}_{I}\!\bra{n;t}\frac{\partial}{\partial t}\ket{n;t}_I = \frac{1}{\sqrt{n}} \, {}_{I}\!\bra{n;t}\frac{\partial \hat{a}^{\dagger}(t)}{\partial t}\ket{n-1;t}_I \, + \, {}_I\!\bra{n-1;t}\frac{\partial}{\partial t}\ket{n-1;t}_{I}.
\end{equation}
As the expression of $\frac{\partial \hat{a}^{\dagger}(t)}{\partial t}$ in terms of $\hat{a}(t)$ and $\hat{a}^{\dagger}(t)$ is (see App.~\ref{app:derivative_of_a})
\begin{equation}\label{adaggerpunto}
\frac{\partial \hat{a}^{\dagger}(t)}{\partial t} = \frac{1}{2}\left\lbrace\left[ \textit{i}M\Big( \sigma(t)\ddot{\sigma}(t)-\dot{\sigma}^2(t) \Big) - 2\frac{\dot{\sigma}(t)}{\sigma(t)} \right]\hat{a}(t) + iM\Big( \sigma(t)\ddot{\sigma}(t)-\dot{\sigma}^2(t) \Big)\hat{a}^{\dagger}(t) \right\rbrace.
\end{equation}
Therefore, from \eqref{diag_element_time-derivative}, we get:
\begin{equation}\label{first_recursion}
{}_I\!\bra{n;t}\frac{\partial}{\partial t}\ket{n;t}_I = {}_{I}\!\bra{n-1;t}\frac{\partial}{\partial t}\ket{n-1;t}_I \, + \frac{iM}{2}\left(\sigma(t)\ddot{\sigma}(t)-\dot{\sigma}^2(t) \right),
\end{equation}
and applying repeatedly \eqref{first_recursion}, we determine that
\begin{equation}\label{DiagonalMatrixElementdt}
{}_I\!\bra{n;t}\frac{\partial}{\partial t}\ket{n;t}_I = {}_{I}\!\bra{0;t}\frac{\partial}{\partial t}\ket{0;t}_I \, +\frac{inM}{2}\left(\sigma(t)\ddot{\sigma}(t)-\dot{\sigma}^2(t) \right).
\end{equation}
In order to complete the calculation of the diagonal matrix elements of the operator $\frac{\partial}{\partial t}$ in the eigenbasis of $\hat{I}$, we need to evaluate ${}_I\!\bra{0;t}\frac{\partial}{\partial t}\ket{0;t}_I $. In this regard, defining the overlap
$\phi_0(q;t)\equiv\bra{q}\ket{0;t}_I$, the quantities ${}_I\!\bra{0;t}\frac{\partial}{\partial t}\ket{0;t}_I$ can be expressed through the following integral:
\begin{equation}\label{phizero_integral}
{}_I\!\bra{0;t}\frac{\partial}{\partial t}\ket{0;t}_I = \int_{-\infty }^{+\infty }dq \,\phi_0^{*}(q;t)\frac{\partial \phi_0(q;t)}{\partial t}.
\end{equation}
The explicit expression of $\phi_0(q;t)$ can be derived by performing the scalar product of $\hat{a}(t)\ket{0;t}_I$ and $\ket{q}$, i.e.,
\begin{equation}\label{eq:scalar_product_q_a0}
\bra{q}\hat{a}(t)\ket{0;t}_I = \frac{1}{\sqrt{2\hbar}}\left( \frac{q}{\sigma(t)} - i\Big( i\hbar \sigma(t) \frac{\partial}{\partial q} + M\dot{\sigma}(t)q\Big) \right) \phi_0(q;t)=0,
\end{equation}
where we used \eqref{cd_definition}, and the equalities $\hat{p}\ket{q}=-i\hbar\frac{\partial}{\partial q}\ket{q}$ and $\hat{a}(t)\ket{0;t}_I=0$. As in \eqref{eq:scalar_product_q_a0} $q$ and $\frac{\partial}{\partial q}$ are functions and not operators, \eqref{eq:scalar_product_q_a0} simplifies as
\begin{equation}\label{eq:scalar_product_q_a0_2}
\left(\frac{1}{\sigma(t)}-iM\dot{\sigma}(t) \right)q \, \phi_0(q;t) + \hbar\sigma(t)\frac{\partial \phi_0(q;t)}{\partial q}=0.
\end{equation}
\eqref{eq:scalar_product_q_a0_2} is a separable differential equation that can be directly integrated, obtaining
\begin{equation}\label{eq:phi_0_q}
\phi_0(q;t)= \mathcal{N} \exp\left( \frac{iM}{\hbar}\left(\frac{\dot{\sigma}(t)}{\sigma(t)}+\frac{i}{M\sigma^2(t)} \right)\frac{q^2}{2} \right), 
\end{equation}
where the normalization $\mathcal{N}=\left(\frac{1}{\pi\hbar\sigma^2(t)} \right)^{1/4}$ is determined by imposing that the 2-norm of $\phi_0(q;t)$ is equal to $1$. In this way, through substitution, the integral in \eqref{phizero_integral} becomes a Gaussian integral that admits the following close-form expression (App.~\ref{app:solution_phizero}): 
\begin{equation}\label{solution_phizero_integral}
{}_I\!\bra{0;t}\frac{\partial}{\partial t}\ket{0;t}_I=\frac{iM}{4}\left(\sigma(t)\ddot{\sigma}(t)-\dot{\sigma}^2(t) \right). 
\end{equation}
Therefore, by inserting \eqref{solution_phizero_integral} in \eqref{DiagonalMatrixElementdt}, we can determine the expression of ${}_{I}\!\bra{n;t}\frac{\partial}{\partial t}\ket{n;t}_I$ as a function of $\sigma(t)$ and its derivatives: 
\begin{equation}\label{average_dt}
{}_I\!\bra{n;t}\frac{\partial}{\partial t}\ket{n;t}_I=\frac{iM}{2}\left(n+\frac{1}{2}\right) \left(\sigma(t)\ddot{\sigma}(t)-\dot{\sigma}^2(t) \right).
\end{equation}

In conclusion, using Eqs.~(\ref{EPequation2}), (\ref{averageH}), (\ref{average_dt}), \eqref{Phase_ode} simplifies to the following differential equation for the phase function $\alpha_n(t)$:
\begin{equation}\label{phaseodefinal}
\frac{d\alpha_n(t)}{dt}=-\left(n+\frac{1}{2}\right) \frac{1}{M\sigma^2(t)},
\end{equation}
from which we obtain:
\begin{equation}
\alpha_n(t)=-\left(n+\frac{1}{2}\right)\int_{0}^{t} d\varsigma\frac{1}{M\sigma^2(\varsigma)}.
\end{equation}
In App.~\ref{app:alternative_derivation_phase_term} we also report an alternative derivation of the dynamical phase $\alpha_n(t)$, which is based on Ref.~\cite{Pedrosa1997}.

\subsection{General solution of the Schr\"{o}dinger equation for a time-dependent QHO}

Using the LR invariant method, the exact solution of the Schr\"{o}dinger equation for the QHO with time-dependent frequency is
\begin{equation}\label{ket_solution}
\ket{\psi(t)}=\sum_nc_ne^{i\alpha_n(t)}\ket{n;t}_{I}=\sum_nc_n\exp\left( -i\left(n+\frac{1}{2} \right)\int_0^t d\varsigma\frac{1}{M\sigma^2(\varsigma)} \right) \ket{n;t}_{I},
\end{equation}
where $c_n = {}_{I}\!\bra{n;0}\ket{\psi(0)}$ and $\ket{n;t}_I $ is given by \eqref{InvariantFockeigenstates}.
Note that, in order to express the solution of $\ket{\psi(t)}$ in the coordinate representation, we have to perform the scalar product of \eqref{ket_solution} with an eigenket $\ket{q}$ of the position operator. In this way, we obtain:
\begin{equation}
\psi(q;t) = \sum_nc_n\exp\left( -i\left(n+\frac{1}{2} \right)\int_0^t d\varsigma\frac{1}{M\sigma^2(\varsigma)} \right)\phi_n(q;t),
\end{equation}
where $\psi(q,t) = \bra{q}\ket{\psi(t)}$ and $\phi_n(q,t)=\bra{q}\ket{n;t}_I$.
Given the definition of annihilation and creation operators in \eqref{cd_definition}, the explicit expression of $\phi_n(q,t)$ may be not trivial~\cite{Ji_1995_01,Ji_1995_02,Pedrosa1997}.

\subsection{Coordinate representation of the invariant's eigenstates}\label{timedependetunitaryformalism}

In this subsection, we present a method based on a time-dependent unitary transformation that allows us to obtain the coordinate representation of the invariant's eigenstates. The idea is to construct a unitary transformation that makes the invariant explicitly time-independent~\cite{Pedrosa1997,FernandezGuasti2003}. As shown in App.~\ref{app:alternative_derivation_phase_term}, this method can also be used to evaluate the phase term.

First, we fix the notation for how states and operators transform under a time-dependent unitary transformation $\hat{V}(t)$. In this regard, if we define the transformed state as $|\tilde{\psi}(t)\rangle = \hat{V}(t)\ket{\psi(t)}$, a given operator $\hat{A}(t)$ transforms as
\begin{equation}
\tilde{\hat{A}}(t)=\hat{V}(t)\hat{A}(t)\hat{V}^{\dagger}(t).
\end{equation}
Then, let us consider the two unitary operators
\begin{equation}\label{Time-dependent unitarymakingS}
\begin{cases}
\displaystyle{\hat{V}_1(t)=\exp\left(-\frac{iM\dot{\sigma}(t)}{2\hbar\sigma(t)}\hat{q}^2 \right)} \\
\displaystyle{\hat{V}_2(t)=\exp\left(\frac{i\log({\sigma}(t))}{2\hbar} \left\lbrace \hat{p}, \hat{q}\right\rbrace \right).}
\end{cases}
\end{equation}
Using the Baker-Campbell-Hausdorff formula
\begin{equation}\label{BCHformula}
{\rm exp}(\hat{A})\hat{B}\,{\rm exp}(-\hat{A}) = \hat{B}+\left[\hat{A},\hat{B} \right] +\frac{1}{2!}\left[\hat{A},\left[ \hat{A}, \hat{B}\right]\right]  + ... + \frac{1}{n!}\left[ \hat{A},\left[ \hat{A},\left[...,\left[ \hat{A},\hat{B}\right] \right] \right]...\right], 
\end{equation}
we get the following transformation laws for $\hat{q}$ and $\hat{p}$:
\begin{equation}\label{eq:transformation_laws}
\begin{cases}
\displaystyle{\hat{V}_1(t)\hat{q}\hat{V}_1^{\dagger}(t)=\hat{q}} \\
\displaystyle{\hat{V}_1(t)\hat{p}\hat{V}_1^{\dagger}(t)=\hat{p}+M\frac{\dot{\sigma}(t)}{\sigma(t)}\hat{q}} \\
\displaystyle{\hat{V}_2(t)\hat{q}\hat{V}_2^{\dagger}(t)=\sigma(t) \hat{q}} \\
\displaystyle{\hat{V}_2(t)\hat{p}\hat{V}_2^{\dagger}(t)=\frac{\hat{p}}{\sigma(t)}.}
\end{cases}
\end{equation}
In this way, defining the unitary operator $\hat{V}(t)=\hat{V}_2(t)\hat{V}_1(t)$ and using relations in (\ref{eq:transformation_laws}), one has that
\begin{equation}
  \tilde{\hat{I}}=\hat{V}(t)\hat{I}(t)\hat{V}^{\dagger}(t)=\frac{1}{2}\left( \hat{p}^2 + \hat{q}^2\right).
\end{equation}

The unitary transformation $\hat{V}(t)$ maps the invariant $\hat{I}(t)$ to the time-independent Hamiltonian of a QHO with unit mass and frequency. Its spectrum and eigenfunctions are well-known~\cite{konishi,Sakurai,messiah2014quantum}. Let us denote the eigenstates of $\tilde{\hat{I}}$ as $\ket{\tilde{n}}_{\tilde{I}}$; thus, the corresponding eigenfunctions are
\begin{equation}
    \braket{q}{\tilde{n}}_{\tilde{I}} = \tilde{\phi}_n(q)=\frac{1}{\sqrt{2^nn!}}\left( \frac{1}{\pi\hbar} \right)^{\frac{1}{4}}\exp\left(-\frac{q^2}{2\hbar}\right)H_n\left(\frac{q}{\sqrt{\hbar}}\right).
\end{equation}
Consequently, using the equality $\ket{n;t}_I=\hat{V}^{\dagger}(t)\ket{\tilde{n}}_{\tilde{I}}$, we obtain the eigenstates of the invariant $\hat{I}(t)$ in the coordinate $q$ representation, that is
\begin{equation}\label{dynamicalstateswithoutphases}
     \braket{q}{\tilde{n},t}_{I} = {\phi}_n(q;t) = \frac{1}{\sqrt{2^nn!}}\left( \frac{1}{\pi\hbar\sigma^2(t)} \right)^{\frac{1}{4}}\exp\left(-\frac{M}{\hbar}\left( \frac{1}{M\sigma^2(t)}-i\frac{\dot{\sigma}(t)}{\sigma(t)}\right)\frac{q^2}{2}\right) H_n\left(\frac{1}{\sqrt{\hbar}}\frac{q}{\sigma(t)}\right),
\end{equation}
where $H_n(x)$ is the n-th Hermite polynomial of $x$. In deriving \eqref{dynamicalstateswithoutphases}, we have used that $\hat{V}(t)\ket{q}$ is still an eigenstate of the position operator $\hat{q}$ with eigenvalue $\frac{q}{\sigma(t)}$, i.e., $\hat{q}\hat{V}(t)\ket{q} = \frac{q}{\sigma(t)}\hat{V}(t)\ket{q}$.
In conclusion, defining 
\begin{equation}
    \Omega(t)=\frac{1}{M\sigma^2(t)}-i\frac{\dot{\sigma}(t)}{\sigma(t)},
\end{equation}
the n-th dynamical eigenstates of the QHO in the $q$ representation is given by
\begin{equation}\label{ExactWaveFunction}
    e^{i\alpha_n(t)}\braket{q}{\tilde{n},t}_{I} = e^{i\alpha_n(t)}{\phi}_n(q;t) = {\psi}_n(q;t)=\frac{1}{\sqrt{2^nn!}}\left( \frac{1}{\pi\hbar\sigma^2(t)} \right)^{\frac{1}{4}}\exp\left(-\frac{M\Omega(t)}{\hbar}\frac{q^2}{2}\right)H_n\left(\frac{1}{\sqrt{\hbar}}\frac{q}{\sigma(t)}\right)e^{i\alpha_n(t)},
\end{equation}
such that ${\psi}(q;t)=\sum_{n}c_{n}{\psi}_n(q;t)$ with $c_n = {}_I\!\bra{n;0}\ket{\psi(0)}$,
in agreement with \eqref{ket_solution} and Refs.~\cite{Ji_1995_01,Ji_1995_02,Pedrosa1997,FernandezGuasti2003,Chen_STA,DelCampo2014,Dabrowski,Buffoni_Gherardini,DittrichReuter2020}. Note that that in Refs.~\cite{Chen_STA,DelCampo2014,Dabrowski,Buffoni_Gherardini,DittrichReuter2020}, a rescaling of the auxiliary function $\sigma$ is done, with unit mass and $\hbar=1$. 

%%%%%%%%%%%%%%%%%%%%%%%%%%%%%%%%%%%%%%%%
\subsection{Generalization to time-dependent mass}

The whole treatment above can be generalized to the case in which the mass of the QHO is also time-dependent. In such a case, the Hamiltonian of the system is
\begin{equation}\label{eq:qhomasstimedependent}
    \hat{H}(t) = \frac{\hat{p}^2}{2M(t)}+\frac{1}{2}M(t)\omega(t)^2\hat{q}^{2}.
\end{equation}
The goal is to obtain the exact dynamical eigenstate for also the time-dependent Hamiltonian (\ref{eq:qhomasstimedependent}). It can be achieved by mapping \eqref{eq:qhomasstimedependent} to a setting with constant mass. In order to do so, we make the time-rescaling
\begin{equation}\label{timerescaling}
    T = \int_{0}^{t}dt'\frac{1}{M(t')}.
\end{equation}
As \eqref{timerescaling} is bijective and thus invertible, $t$ can be expressed as a function of $T$ as $t=t(T)$.
Accordingly, the Schr\"{o}dinger equation (\ref{SE}) becomes
\begin{equation}\label{eq:SE_time-dep-mass}
i\hbar\frac{\partial }{\partial T}\ket{\psi\left( t(T) \right)} = \overline{\hat{H}}(T) \ket{\psi\left( t(T) \right)},
\end{equation}
where
\begin{equation}
    \overline{\hat{H}}(T) = \frac{\hat{p}^2}{2} + \frac{1}{2}\overline{\omega}^2(T)\hat{q}^2
\end{equation}
with $\overline{\omega}^2(T) = m^2(t(T))\,\omega^2(t(T))$.
Hence, the $n$-th dynamical eigenstate ${\psi}_n(q;T)$, solution of the QHO's dynamics (\ref{eq:SE_time-dep-mass}), is given by \eqref{ExactWaveFunction}, provided $t$ is substituted with the time variable $T$:
\begin{equation}
     {\psi}_n(q;T) = \frac{1}{\sqrt{2^nn!}}\left( \frac{1}{\pi\hbar\sigma^2(T)} \right)^{\frac{1}{4}}\exp\left( -\frac{\overline{\Omega}(T)}{\hbar}\frac{q^2}{2}\right)H_n\left(\frac{1}{\sqrt{\hbar}}\frac{q}{\sigma(T)}\right)e^{i\alpha_n(T)},
\end{equation}
where $\overline{\Omega}(T) \equiv \frac{1}{\sigma^2(T)}-i\frac{\sigma'(T)}{\sigma(T)}$. Here, the primed notation denotes the derivative with respect to $T$.
The auxiliary function $\sigma(T)$ satisfies the Ermakov equation (\ref{EPequation2}) with $m=1$:
\begin{equation}\label{ErmakovTvariable}
    \sigma{''}+\overline{\omega}^2(T)\sigma=\frac{1}{\sigma^3},
\end{equation}
while the phase term $\alpha_n(T)$ is given by
$\alpha_n(T) = - \left(n+\frac{1}{2}\right)\int_{0}^{T} d\varsigma\frac{1}{\sigma^2(\varsigma)}$.

At this point, we can perform the back transformation to the original time $t$. Since $\sigma(t)=\sigma(T(t))$, it follows that $\dot{\sigma}(t)=\sigma'(T)\frac{d T}{dt}=\frac{\sigma'(T)}{M(t)}$. Operating in the same way for the second time-derivative and using \eqref{ErmakovTvariable}, we obtain the differential equation that governs the evolution of $\sigma(t)$:
\begin{equation}\label{eq:extendedeErmakov}
\ddot{\sigma} + d(t)\dot{\sigma} + \omega^2(t)\sigma = \frac{1}{M^2(t)\sigma^3},
\end{equation}
where $d(t)$ is a damping term given by $d(t) = \frac{d }{dt}\log(M(t))$. In order to express the dynamical eigenstates in terms of the original time $t$, we need to find the appropriate dependencies of $\overline{\Omega}(t) \equiv \overline{\Omega}(T(t))$ and $\alpha_n(t) \equiv \alpha_n(T(t))$ on $t$. Again, this can be done through the relations $\frac{d T}{dt}=\frac{1}{M(t)}$ and 
$\dot{\sigma}(t)=\sigma'(T)\frac{d T}{dt}=\frac{\sigma'(T)}{M(t)}$.
In particular, we have:
\begin{equation}
    \overline{\Omega}(t) = \frac{1}{\sigma^2(t)}-i\frac{M(t)\dot{\sigma}(t)}{\sigma(t)}
\end{equation}
and
\begin{equation}
    \alpha_n(t) = -\left( n + \frac{1}{2}\right)\int_{0}^{t} d\varsigma \frac{dT}{d\varsigma}\frac{1}{\sigma^2(\varsigma)} = -\left(n+\frac{1}{2}\right)\int_{0}^{t} d\varsigma\frac{1}{M(\varsigma)\sigma^2(\varsigma)}.
\end{equation}
In conclusion, the $n$-th dynamical state of a QHO with time-dependent mass is
\begin{equation}\label{eq:dyn_eigenstate_td_mass}
     {\psi}_n(q;t) = \frac{1}{\sqrt{2^nn!}}\left( \frac{1}{\pi\hbar\sigma^2(t)} \right)^{\frac{1}{4}}\exp\left( -\frac{\overline{\Omega}(t)}
     {\hbar}\frac{q^2}{2} \right)H_n\left(\frac{1}{\sqrt{\hbar}}\frac{q}{\sigma(t)}\right)e^{i\alpha_n(t)}.
\end{equation}
Formally, the expression of \eqref{eq:dyn_eigenstate_td_mass} is the same as the corresponding one for the constant mass case, with the difference that the auxiliary function $\sigma(t)$ satisfies the modified Ermakov equation [\eqref{eq:extendedeErmakov}], which also comprises a damping term.  

%%%%%%%%%%%%%%%%%%%%%%%%%%%%%%%%%%%%%%%%
\section{Bogoliubov transformation and representation in terms of squeezing}
\label{squeezinginterpretationistantaneus}

The exact dynamical states of the time-dependent QHO can be expressed in terms of the eigenstates of the instantaneous Hamiltonian $\hat{H}(t)$, showing that the dynamics of the QHO can be interpreted in terms of squeezing. A glimpse of this result can be obtained by looking at the expression of $\hat{H}(t)$ in \eqref{Hamiltonian_InvariantOperatorsExpression}. In fact, in terms of the annihilation and creation operators of the invariant, the instantaneous Hamiltonian $\hat{H}(t)$ is quadratic. Hence, as shown in App.~\ref{App:SqueezingBogoliubov}, it can be diagonalized by means of a squeezing transformation. In the following, we present the derivation of such result in a more direct way.

At any time $t$, the invariant $\hat{I}(t)$ can be written in terms of annihilation and creation operators:
\begin{equation}
    \hat{I}(t)=\hbar\left( \hat{a}^\dagger(t)\hat{a}(t)+\frac{1}{2}\right)
\end{equation}
with
\begin{equation}
    \begin{cases}
       \displaystyle{\hat{a}(t)= \frac{1}{\sqrt{2\hbar}}\left( \frac{\hat{q}}{\sigma(t)} + i\Big(\sigma(t)\hat{p}-M\dot{\sigma}(t)\hat{q}\Big)\right)} \\
      \displaystyle{\hat{a}^\dagger(t) = \frac{1}{\sqrt{2\hbar}}\left( \frac{\hat{q}}{\sigma(t)} - i\Big( \sigma(t)\hat{p}-M\dot{\sigma}(t)\hat{q}\Big)\right)}.      
    \end{cases}
\end{equation}
The same procedure can be followed for $\hat{H}(t)=\frac{\hat{p}^2}{2m}+\frac{1}{2}m\omega^2(t)\hat{q}^2$ by introducing the operators
\begin{equation}
    \begin{cases}
       \displaystyle{\hat{b}(t)= \frac{1}{\sqrt{2}}\left( f(t) \hat{q} + i\frac{\hat{p}}{\hbar\, f(t)}\right)} \\
       \displaystyle{\hat{b}^\dagger(t) = \frac{1}{\sqrt{2}}\left( f(t) \hat{q}-i\frac{\hat{p}}{\hbar\, f(t)}\right)}        
    \end{cases}
\end{equation}
with $f(t) \equiv \sqrt{\frac{M\omega(t)}{\hbar}}$, such that $\hat{H}(t)=\hbar\omega(t)\left( \hat{b}^\dagger(t)\hat{b}(t)+\frac{1}{2}\right)$. The eigenstates $\ket{n;t}_H$ of $\hat{H}(t)$ constitute the adiabatic basis. With some algebra, we can determine that the two sets of bosonic operators $\left\{ \hat{a}(t),\hat{a}^{\dagger}(t) \right\}$ and $\left\{ \hat{b}(t),\hat{b}^{\dagger}(t) \right\}$ fulfill the following relations: 
\begin{equation}\label{BogoliubovTransf}
    \begin{cases}
        \displaystyle{\hat{a}(t) = u(t)\hat{b}(t)+v(t)\hat{b}^\dagger(t)} \\ \\
        \displaystyle{\hat{a}^\dagger(t) = u^{*}(t)\hat{b}^\dagger(t)+v^{*}(t)\hat{b}(t),}
    \end{cases}
\end{equation}
where
\begin{equation}\label{BogoliubovExplicitCoefficients}
    \begin{cases}
        \displaystyle{ u(t)=\frac{1}{2\sqrt{m\omega(t)}}\left( \frac{1}{\sigma(t)}+m\omega(t)\sigma(t)-im\dot{\sigma}(t)\right)} \\
       \displaystyle{ v(t)=\frac{1}{2\sqrt{m\omega(t)}}\left( \frac{1}{\sigma(t)}-m\omega(t)\sigma(t)-im\dot{\sigma}(t)\right)}.
    \end{cases}
\end{equation}
At any time $t$ the complex coefficients $u(t)$ and $v(t)$ satisfy the constraint
\begin{equation}
    \left| u(t) \right|^2 - \left| v(t) \right|^2 = 1,
\end{equation}
meaning that the relations in \eqref{BogoliubovTransf} are \textit{Bogoliubov canonical transformations}. A theorem due to von Neumann \cite{vonNeumann1931}  asserts that every Bogoliubov canonical transformation can be represented by a unitary operator, such that 
\begin{equation}
    \begin{cases}\label{T_TransformationLaw}
        \displaystyle{\hat{a}(t) = \hat{T}(t) \hat{b}(t) \hat{T}^\dagger(t)} \\ \\  
        \displaystyle{\hat{a}^\dagger(t)=\hat{T}(t) \hat{b}^\dagger(t) \hat{T}^\dagger(t).}
    \end{cases}
\end{equation}
Here, the unitary operator $\hat{T}(t)$ ($\hat{T}(t)\hat{T}^{\dagger}(t)=\hat{T}^{\dagger}(t)\hat{T}(t)=\hat{\mathbb{I}}$ for any $t$) links the vacuum states $\ket{0;t}_H, \ket{0;t}_I$ to each other: $\ket{0;t}_I = \hat{T}(t)\ket{0;t}_H$. Therefore, we can express the eigenstates of the invariant $\hat{I}(t)$ as a function of the eigenstates of the instantaneous Hamiltonian $\hat{H}(t)$:
\begin{eqnarray}\label{eq:T_transformation}
    \ket{n;t}_I &=\frac{\left( \hat{a}^\dagger(t)\right)^n}{\sqrt{n!}}\ket{0;t}_I = \frac{\left( \hat{T}(t)\ \hat{b}^\dagger(t) \ \hat{T}^\dagger(t)\right)^n}{\sqrt{n!}}\ket{0;t}_I \nonumber \\  
    &= \hat{T}(t)\frac{\left( \hat{b}^\dagger(t)\right)^n}{\sqrt{n!}}\hat{T}^\dagger(t)\hat{T}(t)\ket{0;t}_H = \hat{T}(t)\ket{n;t}_{H}.
\end{eqnarray}
As a result, the exact solution of \eqref{ket_solution} is
\begin{equation}\label{exactdynamicalstateadiabaticbasis}
\ket{\psi(t)}=\sum_nc_ne^{i\alpha_n(t)} \hat{T}(t)\ket{n;t}_{H}  \quad \text{with} \quad c_n = {}_I\!\bra{n;0}\ket{\psi(0)}.
\end{equation}

In order to give a physical interpretation of \eqref{exactdynamicalstateadiabaticbasis}, we need to find the explicit expression of the unitary operator $\hat{T}(t)$. For this purpose, we introduce the real parameters $r(t)$, $\chi(t)$ and $\phi(t)$, defined as in the following:
\begin{equation}\label{eq:real_parameters}
    \begin{cases}
        \displaystyle{\cosh(r(t)) = |u(t)|} \\ \\
        \displaystyle{\cos(\chi(t))=\frac{\Re{u(t)}}{|u(t)|}} \\ \\
        \displaystyle{\cos(\phi(t))}=\frac{\Re{\frac{v(t))}{u(t)}}}{\tanh(r(t))}\,.
    \end{cases}
\end{equation}
For convenience, $r(t)$ and $\phi(t)$ can be expressed in terms of the auxiliary function $\sigma(t)$ and its first derivative:
\begin{equation}
\begin{cases}\label{SqueezingParameters_ModulusPhase}
\displaystyle{
r(t)=
\cosh^{-1}\left\{ \frac{1}{2\sqrt{M\omega(t)}} \left( \left( \frac{1}{\sigma(t)} + M\omega(t)\sigma(t) \right)^2+M^2\dot{\sigma}(t)^2 \right)^{\frac{1}{2}} \right\} 
} \\ \\
\displaystyle{
\chi(t)=
\cos^{-1}\left( \frac{1}{2\sqrt{M\omega(t)}\cosh(r(t))}\left(\frac{1}{\sigma(t)}+M\omega(t)\sigma(t)\right)\right)
} \\ \\
\displaystyle{
\phi(t)=
\cos^{-1}\left( \frac{1}{2M\omega(t)} \frac{\sigma(t)^{-2} -M^2\omega^2(t)\sigma(t)^2 + M^2\dot{\sigma}(t)^2}{2\sinh(r(t))\cosh(r(t)) } \right).
}
\end{cases}
\end{equation}
In this way, \eqref{BogoliubovTransf} becomes:
\begin{equation}
\begin{cases}
    \displaystyle{
    \hat{a}(t) = e^{-i\chi(t)}\cosh\left( r(t) \right)\hat{b}(t) + e^{-i\left(\chi(t)-\phi(t)\right)}\sinh\left( r(t) \right)\hat{b}^\dagger(t)}\\ \\    \displaystyle{\hat{a}^\dagger(t) = e^{i\chi(t)}\cosh(r(t))\hat{b}^\dagger(t)+e^{i\left(\chi(t)-\phi(t)\right)}\sinh(r(t))\hat{b}(t).
    }
\end{cases}  
\end{equation}
This entails that the operator $\hat{T}(t)$ can be expressed as
\begin{equation}\label{UnitaryBogoliubovOp}
    \hat{T}(t) = \hat{S}(t,\xi)\hat{R}(t,\chi),
\end{equation}
where $\hat{S}(t,\xi)$ is the squeezing operator~\cite{barnett1997methods} with $\xi(t) = r(t)e^{i\phi(t)}$ (thus, $r(t)$ and $\phi(t)$ are the squeezing modulus and phase, respectively):
\begin{equation}\label{SqueezingOperatorStandard}
    \hat{S}(t,\xi) = \exp\left( -\frac{\xi(t)}{2} \hat{b}^{\dagger2}(t)+\frac{\xi(t)^*}{2}\hat{b}^2(t)\right),
\end{equation}
while $\hat{R}(t,\chi)$ is the phase rotation operator
\begin{equation}
    \hat{R}(t,\chi)=\exp\left( i\chi(t)\left(  \hat{b}^\dagger(t)\hat{b}(t)+\frac{1}{2}\right) \right).
\end{equation}

This result can be derived by considering how $b(t)$ transforms under the action of the operators $\hat{R}(t,\chi)$ and $\hat{S}(t,\xi)$. On the one hand, using \eqref{BCHformula} and the equality $\left[ \hat{b}^\dagger(t)\hat{b(t)},\hat{b}(t) \right]=-\hat{b}(t)$, we get:
\begin{eqnarray}\label{Transf_1}
    \hat{R}(t,\chi) \hat{b}(t) \hat{R}^\dagger(t,\chi) &=& 
    \displaystyle{
    \hat{b}(t)+i\chi(t)\left[ \hat{b}^\dagger(t)\hat{b}(t),\hat{b}(t)) \right] + ... 
    + \frac{(i\chi(t))^n}{n!}\left[ \hat{b}^\dagger(t)\hat{b}(t),\left[ \hat{b}^\dagger(t)\hat{b}(t),\left[...,\left[ \hat{b}^\dagger(t)\hat{b}(t),\hat{b}(t)\right] \right] \right]...\right]
    }\nonumber \\ 
    &=& \displaystyle{\hat{b}(t)-i\chi(t)\hat{b}(t)+\frac{(i\chi(t))^2}{2!}\hat{b}(t)-\frac{(i\chi(t))^3}{3!}\hat{b}(t)+\ ...\ \frac{(-i\chi(t))^n}{n!}\hat{b}(t)} \nonumber \\ 
    &=& \displaystyle{\sum_{n=0}^{+\infty }\frac{\left( -i\chi\left( t \right) \right)^n}{n!}\ \hat{b}(t)=e^{-i\chi(t)}\hat{b}(t).}
\end{eqnarray}
The transformation law for $\hat{b}^\dagger(t)$ is then determined by taking the adjoint of both sides of \eqref{Transf_1}:
\begin{equation}\label{Transf_2}
\hat{R}(t,\chi) \hat{b}^\dagger(t) \hat{R}^\dagger(t,\chi) = e^{i\chi(t)}\hat{b}^\dagger(t).
\end{equation}
On the other hand, the action of the squeezing operator $\hat{S}(t,\xi)$ on the annihilation and creation operators $\hat{b}(t), \hat{b}^\dagger(t)$ is a standard results of quantum optics~\cite{barnett1997methods,Varro_2022}. It is achieved by exploiting the form of $\hat{S}(t,\xi)$, the Baker-Campbell-Hausdorff formula [\eqref{BCHformula}] and the algebra of the QHO:
\begin{eqnarray}
\label{Transf_3}
\hat{S}(t,\xi) \hat{b}(t) \hat{S}^\dagger(t,\xi)
&=&\hat{b}(t)+\xi(t)\hat{b}^\dagger(t)+\frac{1}{2!}\left| \xi(t) \right|^2\hat{b}(t)+\frac{1}{3!}\xi(t)|\xi(t)|^2\hat{b}^\dagger(t)+\ ...\nonumber \\ 
&=&\cosh(r(t))\hat{b}(t)+e^{i\phi(t)}\sinh(r(t))\hat{b}^\dagger(t).
\end{eqnarray}
Therefore, making the Hermitian conjugate of \eqref{Transf_3}, we find:
\begin{equation}\label{Transf_4}
    \hat{S}(t,\xi) \hat{b}^\dagger(t) \hat{S}^\dagger(t,\xi)= \cosh(r(t))\hat{b}^{\dagger}(t)+e^{-i\phi(t)}\sinh(r(t))\hat{b}(t).
\end{equation}
Finally, with the aid of Eqs.~(\ref{Transf_1}), (\ref{Transf_2}), (\ref{Transf_3}) and (\ref{Transf_4}) it can be proved that $\hat{T}(t)$ is the unitary operator associated to Eqs.~(\ref{BogoliubovTransf})-(\ref{BogoliubovExplicitCoefficients}).

If the system is initially in the $n$-th instantaneous eigenstate $\ket{n;t}_I = \hat{T}(t)\ket{n;t}_H$ [\eqref{eq:T_transformation}] of the invariant $\hat{I}(t)$, then the exact evolved state of the driven QHO is
\begin{equation}
\ket{\psi_n(t)}=e^{i\alpha_n(t)} \hat{T}(t)\ket{n;t}_H=e^{i\alpha_n(t)}e^{i(n+\frac{1}{2})\chi(t)}\hat{S}(t,\xi)\ket{n;t}_{H},
\end{equation}
which can be thus interpreted (apart from time-dependent phases) as a {\it squeezed-number state} of $\ket{n;t}_H$, instantaneous eigenstate of $\hat{H}(t)$. In conclusion, using Eqs.~(\ref{eq:InvariantEvolutionOperator}), (\ref{exactdynamicalstateadiabaticbasis}) and (\ref{UnitaryBogoliubovOp}), we determine the evolution operator
\begin{equation}\label{eq:QHOEvolutionOperator}
    \hat{U}(t,0)=\sum_{n}^{}e^{i\alpha_n(t)}e^{i(n+\frac{1}{2})\chi(t)}\hat{S}(t,\xi)\ket{n;t}_{H}\!\bra{n;0}.
\end{equation}

\subsection{Variance of $\hat{q}$ and $\hat{p}$ with respect to $\ket{\psi_n(t)}$}

The variance of position and momentum operators with respect to the $n$-th dynamical state $\ket{\psi_n(t)}$ can be evaluated through Eqs.~(\ref{BogoliubovTransf})-(\ref{BogoliubovExplicitCoefficients}). Here, we show the main steps of the calculations that lead to the expression of the variance $\Delta\hat q_n^2(t)$ of $\hat{q}$. We start from computing $\left\langle \hat{q} \right\rangle_n(t) \equiv \bra{\psi_n(t)}\hat{q}\ket{\psi_n(t)}$:
\begin{eqnarray}\label{eq:averageposition}
        \left\langle \hat{q} \right\rangle_n(t) &=& \frac{1}{\sqrt{2}f(t)} \bra{\psi_n(t)}\left( \hat{b}(t) + \hat{b}^{\dagger}(t)\right) \ket{\psi_n(t)}\nonumber \\
        &=& \frac{1}{\sqrt{2}f(t)} {}_{I}\!\bra{n;t}\Big[ \left( u(t)^* -v(t)^* \right)\hat{a}(t) + \left( u(t) - v(t) \right)\hat{a}^\dagger(t)\Big]\ket{n;t}_{I} = 0 \quad \text{for any} \,\,t.
\end{eqnarray}
As a consequence, $\Delta\hat q_n^2(t) = \left\langle \hat{q}^2\right\rangle_n(t) \equiv \bra{\psi_n(t)}\hat{q}^2\ket{\psi_n(t)}$ that reads as
\begin{eqnarray}\label{eq:average squared position}
    \left\langle \hat{q}^2\right\rangle_n(t) &=& \frac{1}{2f(t)^2} \bra{\psi_n(t)}\left( \hat{b}(t)+\hat{b}^{\dagger}(t)\right)^2\ket{\psi_n(t)} \nonumber \\
    &=& \frac{1}{2f(t)^2} {}_{I}\!\bra{n;t}\Big[ \big( u(t)^*-v(t)^* \big)\hat{a}(t) + \big( u(t) -v(t)\big)\hat{a}^\dagger(t) \Big]^2\ket{n;t}_{I} \nonumber \\
    &=& \frac{1}{2f(t)^2}\left|u(t) -v(t)\right|^2{}_{I}\!\bra{n;t}\Big( \hat{a}(t)\hat{a}^\dagger(t)+\hat{a}^\dagger(t)\hat{a}(t)\Big) \ket{n;t}_{I} = \frac{\hbar}{M\omega(t)}\left(n+\frac{1}{2}\right)\left|u(t)-v(t)\right|^{2}.
\end{eqnarray}
Notice that the results of Eqs.~(\ref{eq:averageposition})-(\ref{eq:average squared position}) are obtained using the algebra of the time-dependent creation and annihilation ladder operators. After similar calculations for $\hat{p}$, we finally obtain:
\begin{equation}
\begin{cases}\label{position-momentumvariance}
    \Delta\hat q_n^2(t) = \frac{\hbar}{M\omega(t)}\left(n+\frac{1}{2}\right)\left|u(t) - v(t)\right|^2=\hbar\sigma^2(t)\left(n+\frac{1}{2}\right)\\ \\ 
    \Delta\hat p_n^2(t)=\hbar M\omega(t)\left(n+\frac{1}{2}\right)\left|u(t) + v(t)\right|^2=\hbar\left( \frac{1}{\sigma^2(t)}+M\dot{\sigma}^2(t)\right) \left(n+\frac{1}{2}\right).
\end{cases}
\end{equation}
Moreover, using $u(t)=e^{-i\chi(t)}\cosh(r(t))$ and $v(t) = e^{-i(\chi(t)-\phi(t))}\sinh(r(t))$, \eqref{position-momentumvariance} can be expressed as a function of the squeezing modulus and phase:
\begin{equation}
\begin{cases}
    \Delta\hat q_n^2(t)=\frac{\hbar}{M\omega(t)}\left(n+\frac{1}{2}\right)\Big( \cosh^2(r(t))+\sinh^2(r(t))-2\sinh(r(t))\cosh(r(t))\cos(\phi(t)) \Big)\\ \\
    \Delta\hat p_n^2(t)=\hbar M\omega(t)\left(n+\frac{1}{2}\right)\Big( \cosh^2(r(t))+\sinh^2(r(t))+2\sinh(r(t))\cosh(r(t))\cos(\phi(t)) \Big).
\end{cases}
\end{equation}
Notice that \eqref{position-momentumvariance} can also be evaluated starting from the expression of $\hat{q}, \hat{p}, \hat{q}^2, \hat{p}^2$ reported in App.~\ref{app:definitions_L}.

\subsection{Application to coherent states}

In this section, we study the eigenstates of the time-dependent annihilation operator $\hat{a}(t)$ of the invariant, using the 
Lewis-Riesenfeld theory. We will show how these eigenstates can be interpreted as th  coherent states for the time-dependent QHO~\cite{CoherentStatesTDQH_Ray,Pedrosa_1987_SqueezedCoherntStates}. 

At the initial time $t=0$ the QHO is in a state of the form $\ket{\alpha,0}=\sum_{n}^{}c_n\ket{n;0}_I$ with $\alpha\in\mathbb{C}$. We require $\ket{\alpha;0}_I$ to be an eigenstate of $\hat{a}(0)$ with eigenvalue $\alpha$: $\hat{a}(0)\ket{\alpha,0}=\alpha\ket{\alpha,0}$. With standard manipulations \cite{barnett1997methods}, we find that
\begin{equation}
    \ket{\alpha,0}=e^{-\frac{\left|\alpha\right|^2}{2}}\sum_{n=0}^{+\infty}\frac{\alpha^n}{\sqrt{n!}}\ket{n;0}_{I}.
\end{equation}
The evolved state at the generic instant $t$ can be determined using \eqref{ket_solution}:
\begin{equation}\label{CoherentStaseTDQHO}
    \ket{\alpha,t}=e^{-\frac{\left|\alpha\right|^2}{2}}\sum_{n=0}^{+\infty}\frac{\alpha^n}{\sqrt{n!}}e^{i\alpha_n(t)}\ket{n;t}_{I},
\end{equation}
where, we recall, $\alpha_n\left(t\right)=-\left(n+\frac{1}{2}\right)\int_{0}^{t}dt'\frac{1}{M\sigma^2(t')}$.
Let us apply $\hat{a}(t)$ [see \eqref{cd_definition}] to the evolved state $\ket{\alpha,t}$; we get:
\begin{equation}
\begin{split}
    \hat{a}(t)\ket{\alpha,t} &= e^{-\frac{\left|\alpha\right|^2}{2}}\sum_{n=0}^{+\infty}\frac{\alpha^n}{\sqrt{n!}}e^{i\alpha_n(t)}\hat{a}(t)\ket{n;t}_I = e^{-\frac{\left|\alpha\right|^2}{2}}\sum_{n=1}^{+\infty}\frac{\alpha^n}{\sqrt{n!}}\sqrt{n} \, e^{i\alpha_n(t)}\ket{n-1;t}_I \\
    &= e^{-\frac{\left|\alpha\right|^2}{2}}\sum_{n=0}^{+\infty}\frac{\alpha^{n+1}}{\sqrt{n!}}e^{i\alpha_{n+1}(t)}\ket{n;t}_I = \alpha \, \exp\left( -i\int_{t_0}^{t}dt'\frac{1}{m\sigma^2(t')}\right) \ket{\alpha,t} = \alpha \, e^{2i\alpha_0(t)}\ket{\alpha,t}.
\end{split}
\end{equation}
Hence, the state $\ket{\alpha,t}$ [\eqref{CoherentStaseTDQHO}] is an eigenstate of $\hat{a}(t)$ and, at the same time, is the evolved state of $\ket{\alpha,0}$. It thus represents the coherent state for the time-dependent QHO.

In order to provide physical interpretation of the results by means of the squeezing operator, we repeat the treatment in Sec.~\ref{squeezinginterpretationistantaneus} using the creation and annihilation operators of the initial Hamiltonian $\hat{H}(0)$, i.e., $\hat{b}(0)= \frac{1}{\sqrt{2}}\left( f(0) \hat{q} + i\frac{\hat{p}}{\hbar\,f(0)}\right)$ and $\hat{b}^\dagger(0) = \frac{1}{\sqrt{2}}\left( f(0) \hat{q}-i\frac{\hat{p}}{\hbar\,f(0)}\right)$, such that $\hat{H}(0)=\hbar\omega(0)\left( \hat{b}^\dagger(0)\hat{b}(0)+\frac{1}{2} \right)$. The operators $\hat{a}(t)$ and $\hat{a}^\dagger(t)$ can be expressed as 
\begin{equation}\label{BogoliubovTransfInitialBasis}
    \begin{cases}
        \displaystyle{\hat{a}(t) = \mu(t)\hat{b}(0)+\nu(t)\hat{b}^\dagger(0)} \\ \\
        \displaystyle{\hat{a}^\dagger(t) = \mu^{*}(t)\hat{b}^\dagger(0)+\nu^{*}(t)\hat{b}(0),}
    \end{cases}
\end{equation}
where
\begin{equation}
    \begin{cases}
        \displaystyle{ 
        \mu(t)=\frac{1}{2\sqrt{M\omega(0)}}\left( \frac{1}{\sigma(t)}+M\omega(0)\sigma(t)-iM\dot{\sigma}(t)\right)
        } \\
       \displaystyle{ 
       \nu(t)=\frac{1}{2\sqrt{M\omega(0)}}\left( \frac{1}{\sigma(t)}-M\omega(0)\sigma(t)-iM\dot{\sigma}(t)\right)
       }
    \end{cases}
\end{equation}
with the complex coefficients $\mu(t)$ and $\nu(t)$ satisfying the relation $\left|\mu(t)\right|^2-\left|\nu(t)\right|^2=1$. Therefore, the coherent states $\ket{\alpha,t}$ of \eqref{CoherentStaseTDQHO} can be interpreted as the \textit{squeezed coherent states}, also known as the \textit{two-photon coherent state}~\cite{Yuen1976}, of quantum optics~\cite{barnett1997methods}. Thanks to this identification, we can straightforwardly evaluate the uncertainty of $\hat{q}$ and $\hat{p}$ for the states $\ket{\alpha,t}$:
\begin{equation}
\begin{cases}\label{position-momentumvariancecoherentstatetdqho}
    \Delta\hat q_{\alpha}^2(t) = \frac{\hbar}{2M\omega(0)}\left|\mu(t) -\nu(t) \right|^2 = \frac{\hbar\sigma^2(t)}{2}\\ \\ 
    \Delta\hat p_{\alpha}^2(t)=\frac{\hbar M\omega(0)}{2}\left|\mu(t) + \nu(t) \right|^2 = \frac{\hbar}{2}\left( \frac{1}{\sigma^2(t)} + M\dot{\sigma}^2(t)\right).
\end{cases}
\end{equation}
Moreover, the motion of the wave-packet described by $\ket{\alpha,t}$ is given by the expectation value of $\hat{q}$ with respect to $\ket{\alpha,t}$: 
\begin{equation}\label{averagepositionCoherentstate}
    \bra{\alpha,t}\hat{q}\ket{\alpha,t} = \sqrt{\frac{\hbar}{2}}\sigma(t)\bra{\alpha,t}\left[\hat{a}(t)+\hat{a}^\dagger(t)\right]\ket{\alpha,t} = \sqrt{ 2\hbar }\, \sigma(t)\left|\alpha\right|\cos\left( \Lambda(t) + \theta \right),
\end{equation}
where $\Lambda(t) \equiv -2\alpha_0(t)$ and $\alpha=\left|\alpha\right|e^{-i\theta}$. \eqref{averagepositionCoherentstate} has the same form of the solution for the dynamics of a classical harmonic oscillator~\cite{RAY19804,Ray1982}. We can thus conclude that averaging over coherent states reproduces the classical motion.

%%%%%%%%%%%%%%%%%%%%%%%%%%%%%%%%%%%%%%%%
\section{Excitation amplitudes with initial superposition states}\label{Section:Exitationampl}

In this section, we discuss how to determine the expression of the amplitudes ${}_{H}\!\bra{m;t}\ket{\psi(t)}$, whose modulus squared gives the probabilities that at a generic time $t$ the QHO is in an eigenstate $\ket{m;t}_H$ of $\hat{H}(t)$. We consider that the system is initialized at $t=0$ in a superposition state $|\psi(0)\rangle = \sum_{k}g_{k}\ket{k;0}_H$. Notice that the latter is the most general initial pure state that can be considered, under the assumption that the eigenstates of $\hat{H}(t)$ form a complete basis of the system's Hilbert space at any time $t$.

At the initial time we impose that the invariant is directly proportional to the Hamiltonian: $\hat{I}(0)\propto \hat{H}(0)=\frac{\hat{p}^2}{2M}+\frac{1}{2}M\omega(0)^2\hat{q}^2$. This is valid by taking $\sigma(0) = \frac{1}{\sqrt{M\omega(0)}}$ and $\dot{\sigma}(0) = 0$ as initial conditions of the Ermakov equation. Here, this choice of initial conditions may seem completely arbitrary; a justification of it will be given later.

In the following, we are going to present two different methods to derive the amplitudes ${}_{H}\!\bra{m;t}\ket{\psi(t)}$. The first approach is algebraic and based on interpreting the dynamics of a driven QHO as a squeezing transformation with respect to the eigenbasis of the instantaneous Hamiltonian. The second method, reported in App.\ref{APP: AnalyticMethod}, directly considers the exact dynamical wave-function solving the Schr\"{o}dinger equation in the coordinate representation. It is worth keeping in mind the difference between ${\psi}_n(q;t) = e^{i\alpha_n(t)}\braket{q}{\tilde{n},t}_{I}$, given by \eqref{ExactWaveFunction}, and the wave-function 
\begin{equation}
     \braket{q}{n,t}_{H} \equiv {\Phi}_n(q;t) = \frac{1}{\sqrt{2^nn!}}\left( \frac{M\omega(t)}{\hbar} \frac{1}{\pi}\right)^{\frac{1}{4}}\exp\left( -\frac{M\omega(t)}{\hbar}\frac{q^2}{2} \right)H_n\left( \sqrt{\frac{M\omega(t)}{\hbar}}q \right)
\end{equation}
associated to the eigenstate $\ket{n;t}_H$ of $\hat{H}(t)$. Such a difference will become clearer below.

\subsection{Algebraic method}

In Sec.~\ref{squeezinginterpretationistantaneus} we have derived that the exact dynamics of a driven QHO can be described as a squeezing transformation of the adiabatic bases. More precisely, the dynamical state of the QHO at time $t$ is
\begin{equation}\label{Oscillator_final_State}
\ket{\psi(t)} = \sum_{n}c_{n}e^{i\alpha_n(t)}\hat{T}(t)\ket{n;t}_H
=\sum_{n}c_{n}e^{i\alpha_n(t)}e^{i(n+\frac{1}{2})\chi(t)}\hat{S}(t,\xi)\ket{n;t}_{H}.
\end{equation}
As a result, the amplitude ${}_{H}\!\bra{m;t}\ket{\psi(t)}$ can be written as
\begin{equation}\label{AmplitudesSqueezing}    {}_{H}\!\bra{m;t}\ket{\psi(t)} = \sum_{n}c_ne^{i\alpha_n(t)}e^{i\left(n+\frac{1}{2}  \right)\chi(t)}{}_{H}\!\bra{m;t}\hat{S}(t,\xi)\ket{n;t}_{H}.
\end{equation}
We recall that $c_n \equiv {}_{I}\!\bra{n;0}\ket{\psi(0)}$; thus, if $\ket{\psi(0)}=\sum_{k}g_{k}\ket{k;0}_{H}$, then
\begin{equation}
    c_n = \sum_{k} g_{k} \, {}_{H}\!\bra{n;0}\hat{T}^\dagger(0)\ket{k;0}_{H}.
\end{equation}
The operator $\hat{T}(0)$ depends on the initial conditions $\sigma(0)$, $\dot{\sigma}(0)$ of the Ermakov equation. The coefficient ${}_{H}\!\bra{n;0}\hat{T}^\dagger(0)\ket{k;0}_{H}$ is, by construction, the component of the $k$-th eigenstate of $\hat{H}(0)$ with respect to the $n$-th eigenstate of $\hat{I}(0)$. The ideal choice is thus to take $\hat{I}(0)\propto\hat{H}(0)$, such that $\hat{T}(0)=\hat{\mathbb{I}}$ and
\begin{equation}
c_n=\sum_{k}g_{k} \, {}_{H}\!\bra{n;0}\ket{k;0}_{H} = g_{n}.
\end{equation}
Hence, through \eqref{AmplitudesSqueezing}, the calculation of the amplitudes ${}_{H}\!\bra{m;t}\ket{\psi(t)}$ is carried out by computing the matrix elements of the squeezing operator between Fock states.

Before determining the expression of ${}_{H}\!\bra{m;t}\ket{\psi(t)}$, let us consider a special case already discussed in the current literature, namely that at the initial time $t=0$ the QHO is in the ground state of $\hat{H}(0)$. In such a case, the amplitude (\ref{AmplitudesSqueezing}) simplifies to ${}_{H}\!\bra{m;t}\ket{\psi(t)}=e^{i\alpha_0(t)}e^{i\frac{\chi(t)}{2}} {}_{H}\!\bra{m;t}\hat{S}(t,\xi)\ket{0;t}_H$ with $m$ a generic integer. To evaluate the action the squeezing operator $\hat{S}(t,\xi)$ of \eqref{SqueezingOperatorStandard} on the ground state of $\hat{H}(0)$, we express it in the normal ordered form \cite{barnett1997methods}:
\begin{eqnarray}\label{OrderedSqueezingOperator}
    \hat{S}(t,\xi) &=& \exp\left(-\frac{1}{2}\hat{b}^{\dagger}(t)^2 e^{i\phi(t)}\tanh(r(t))\right)
    \exp\left(-\frac{1}{2}\left(\hat{b}^\dagger(t) \hat{b}(t) + \hat{b}(t)\hat{b}^\dagger(t)\right)\log\cosh(r(t)) \right)\nonumber \\
    &&\times\exp\left( \frac{1}{2}\hat{b}^2(t) e^{-i\phi(t)}\tanh(r(t)) \right).
\end{eqnarray}
Remembering that the exponential of an operator is defined by the corresponding power series and the action of the annihilation operator $\hat{b}(t)$ on $\ket{0;t}_H$ is $\hat{b}\ket{0;t}_H=0$, we get:
\begin{equation}\label{eq:S_applied}
    \hat{S}(t,\xi)\ket{0;t}_H = \frac{1}{\sqrt{\cosh(r(t))}}\sum_{k=0}^{\infty }\frac{\sqrt{(2k)!}}{k!}\left( -\frac{1}{2}e^{i\phi(t)}\tanh\left( r(t) \right) \right)^{k}\ket{2k;t}_{H}.
\end{equation}
From \eqref{eq:S_applied}, we determine that $m$ must be even, as expected from the selection rules due to parity. As a result,
\begin{equation}
\begin{split}
    {}_{H}\!\bra{2k;t}\ket{\psi(t)} &= e^{i\alpha_0(t)}e^{\frac{i\chi(t)}{2}}{}_{H}\!\bra{2k;t}\hat{S}(t,\xi)\ket{0;t}_H \\ 
    &= e^{i\alpha_0(t)}e^{\frac{i\chi(t)}{2}}\frac{1}{\sqrt{\cosh(r(t))}}\frac{\sqrt{2k!}}{k!}\left( -\frac{1}{2}e^{i\phi(t)}\tanh\big(r(t)\big)\right)^k
\end{split}
\end{equation}
with $k=0,1,\ldots$. Therefore, the probability that a driven QHO, initialized in the ground state $\hat{H}(0)$, develops $2k$ excitations at time $t$ is
\begin{equation}\label{ExitationProbGround}
    p(2k,t) = \Big\vert {}_{H}\!\bra{2k;t}\ket{\psi(t)} \Big\vert^2 = \frac{(2k)!}{(k!)^22^{2k}}\frac{1}{\cosh\big( r(t) \big)}\Big( \tanh\big( r(t) \big)\Big)^{2k}.
\end{equation}
This expression is equivalent to Eq.~(46) of the supplementary material of Ref.~\cite{Buffoni_Gherardini}.
In order to show this, using the formalism discussed above, we consider again the Bogoliubov coefficients $u(t) = \frac{\sigma(t)}{2\sqrt{M\omega(t)}}\left( \frac{1}{\sigma^2(t)}+M\omega(t)-iM\frac{\dot{\sigma}(t)}{\sigma(t)} \right)$ and $v(t) = \frac{\sigma(t)}{2\sqrt{M\omega(t)}}\left( \frac{1}{\sigma^2(t)}-M\omega(t)+iM\frac{\dot{\sigma}(t)}{\sigma(t)} \right)$, such that
\begin{equation}\label{eq:tanh-r_square}
    \Big( \tanh\big( r(t) \big)\Big)^2 = \frac{|v(t)|^2}{|u(t)|^2} = \frac{ \left( \frac{1}{\sigma^2(t)}-M\omega(t)\right)^2 +M^2 \frac{\dot{\sigma}^2(t)}{\sigma^2(t)}}{\left( \frac{1}{\sigma^2(t)}+M\omega(t)\right)^2 + M^2\frac{\dot{\sigma}^2(t)}{\sigma^2(t)}}.
\end{equation}
\eqref{eq:tanh-r_square} corresponds to $|R_t|^2$ defined in Ref.~\cite{Buffoni_Gherardini}, in which $\xi(t)$ is related to $\sigma(t)$ through the relation $\xi(t)=\sigma(t)/\sqrt{2}$ as $M=1$. In particular, using the definitions of hyperbolic functions, we have: $\frac{1}{\cosh(r(t))} = \sqrt{1-(\tanh(r(t)))^2} = \sqrt{1-|R_t|^2}$. Hence, if we set $2k=m$ in \eqref{ExitationProbGround} and we use $2^mm!=(2m)!!$, $m!=m!!(m-1)!!$, we get:
\begin{equation}\label{probampl_Rfactor}
    p(m,t) = \frac{(m-1)!!}{m!!}\sqrt{1-|R_t|^2} \, |R_t|^{m},
\end{equation}
in exact agreement with Ref.~\cite{Buffoni_Gherardini}. Finally, introducing the parameter $\tilde{p}(t) = \frac{1}{\cosh^2{(r(t))}}$, \eqref{ExitationProbGround} simplifies to
\begin{equation}\label{eq:NegativeBinomial}
    p(2k,t) = \binom{k-\frac{1}{2}}{k} \tilde{p}(t)^\frac{1}{2}\left( 1-\tilde{p}(t) \right)^{k}
\end{equation}
that is a negative binomial distribution~\cite{Buffoni_Gherardini,GomezRuizQST2025}. \eqref{eq:NegativeBinomial} is valid for any protocol that varies the frequency of the QHO over time. This time-dependence, indeed, is fully encoded in $\tilde{p}(t)$ through the squeezing parameter $r(t)$, which is expressed as a function of the Ermakov equation's solution.

%%%%%%%%%%%%%%%%%%%%%%%%%%%%%%%%%%%%%%%%%%%%%%%%%
% Figure 1 - Negative binomial distribution
%%%%%%%%%%%%%%%%%%%%%%%%%%%%%%%%%%%%%%%%%%%%%%%%%
\begin{figure}
    \centering
    \includegraphics[width=0.5\linewidth]{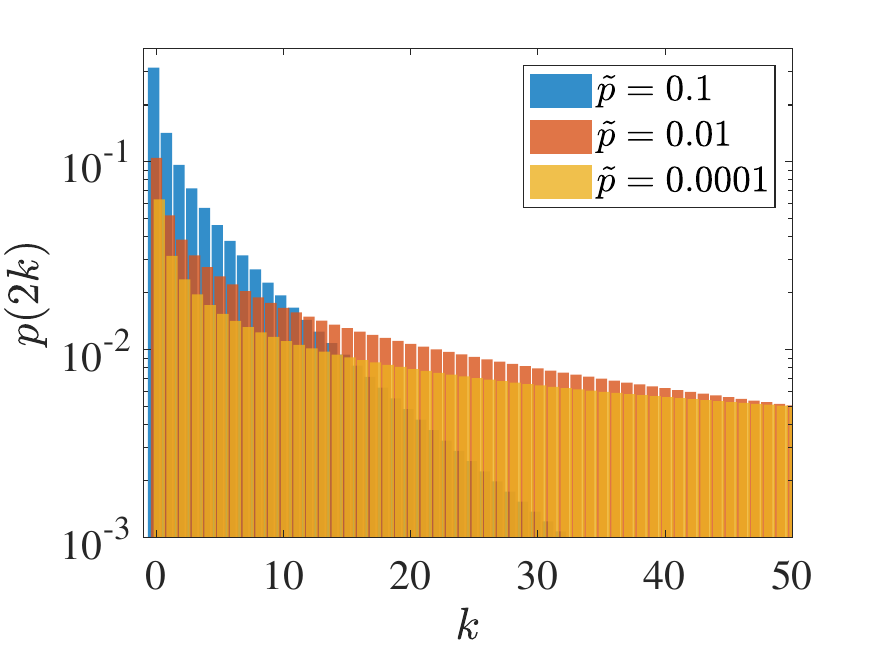}
    \caption{
    Negative binomial distribution as a function of $k$, for $\tilde{p} = 10^{-1}$ (blue histogram), $10^{-2}$ (red histogram) and $10^{-4}$ (yellow histogram). 
    }    
    \label{fig:placeholder}
\end{figure}
%%%%%%%%%%%%%%%%%%%%%%%%%%%%%%%%

Now we address the more general case in which the QHO is in a coherent superposition of eigenstates of $\hat{H}(0)$ at the initial time $t_0=0$. The aim is to determine the amplitude of \eqref{AmplitudesSqueezing}. We thus evaluate the matrix elements of $\hat{S}(t,\xi)$ between the Fock states of $\hat{H}(t)$. Using \eqref{OrderedSqueezingOperator} and the properties of the creation and annihilation operators, we get:
\begin{equation}\label{SquuezingMatrixelementSeries}
    \begin{cases}
        \displaystyle{ {}_{H}\!\bra{m;t}\hat{S}(t,\xi)\ket{n;t}_H = } \\
        \displaystyle{ \exp\left( -\frac{1}{2}\left( n+\frac{1}{2}\right)\eta(t)\right) \left( -\frac{ \zeta(t) }{2}\right)^\frac{m-n}{2}\sqrt{m!n!}\sum_{k=0}^{\left[ n/2 \right]}\frac{\left( -\frac{1}{4}\left| \zeta(t) \right|^2 e^{\eta(t)} \right)^k}{\left[ \frac{1}{2}(m-n)+k \right]! \, k! \, (n-2k)!} \quad (m\ge n);}
        \\
        \displaystyle{ {}_{H}\!\bra{m;t}\hat{S}(t,\xi)\ket{n;t}_H = } \\
        \displaystyle{ \exp\left( -\frac{1}{2} \left( m+\frac{1}{2} \right)\eta(t)\right)\left( \frac{ \zeta^*(t) }{2}\right)^\frac{n-m}{2}\sqrt{m!n!}\sum_{\ell=0}^{\left[ m/2 \right]}\frac{\left( -\frac{1}{4}\left| \zeta(t) \right|^2 e^{\eta(t)} \right)^l}{\left[ \frac{1}{2}(n-m)+\ell \right]! \, \ell! \, (m-2\ell)!} \quad (m\le n),}
    \end{cases}
\end{equation}
with $\eta = 2\log\cosh(r(t))$ and $\zeta(t) = e^{i\phi(t)}\tanh(r(t))$. Notice that if $m, n$ do not have the same parity, then the matrix elements of $\hat{S}(t,\xi)$ are equal to zero. Moreover, the expressions in \eqref{SquuezingMatrixelementSeries} can be rewritten in terms of the Gauss 
hypergeometric function $F(\alpha, \beta; \gamma; z)$~\cite{abramowitz1972handbook}, such that
\begin{equation}\label{SqueezingFockHyper}
    \begin{cases}
        \displaystyle{ {}_{H}\!\bra{m;t}\hat{S}(t,\xi)\ket{n;t}_H = } \\
        \displaystyle{ \: \exp\left( -\frac{1}{2}\left( n+\frac{1}{2}\right)\eta(t)\right)\left( -\frac{\zeta(t) }{2}\right)^\frac{m-n}{2}\sqrt{\frac{m!}{n!}}\frac{1}{\Gamma(\alpha+1)}F \left( -\frac{n}{2},\frac{1-n}{2},\frac{m-n}{2}+1; z(t) \right) \quad (m\ge n);} \\
        \displaystyle{ {}_{H}\!\bra{m;t}\hat{S}(t,\xi)\ket{n;t}_H = } \\
        \displaystyle{ \: \exp\left( -\frac{1}{2}\left( m+\frac{1}{2}\right)\eta(t)\right)\left( \frac{\zeta^*(t) }{2}\right)^\frac{n-m}{2}\sqrt{\frac{n!}{m!}}\frac{1}{\Gamma(\lambda+1)}F\left(-\frac{m}{2},\frac{1-m}{2},\frac{n-m}{2}+1;z(t)\right) \quad (m\le n),}
    \end{cases}
\end{equation}
with $z(t) = -\sinh(r(t))^2$, $\alpha=\frac{1}{2}(m-n)$ and $\lambda=\frac{1}{2}(n-m)$.
As shown in \cite{Varro_2022}, different forms of the matrix elements of $\hat{S}(t,\xi)$ between Fock states can be derived from \eqref{SqueezingFockHyper} using the relations between the Gauss hypergeometric functions and classical orthogonal polynomials~\cite{abramowitz1972handbook}. A compact expression can be obtained in terms of the associated Legendre functions $\mathcal{P}^{\nu}_\mu(x)$:
\begin{equation}\label{SqueezingFockLegendre}
  \begin{cases}
      \displaystyle{ {}_{H}\!\bra{m;t}\hat{S}(t,\xi)\ket{n;t}_H = \sqrt{\frac{n!}{m!}}\sqrt{x(t)}\ \mathcal{P}^{k}_l \left( x(t) \right) \, e^{ik\phi(t)} \quad\quad\quad\quad\quad k=\frac{1}{2}(m-n)\quad\quad (m\ge n);} \\ 
      \displaystyle{ {}_{H}\!\bra{m;t}\hat{S}(t,\xi)\ket{n;t}_H = (-1)^{k'}\sqrt{\frac{m!}{n!}}\sqrt{x(t)}\ \mathcal{P}^{k'}_l \left( x(t) \right) \, e^{-ik'\phi(t)}\quad k'=\frac{1}{2}(n-m)=-k\quad\quad (m\le n),}
    \end{cases}
\end{equation}
where $x(t) = \frac{1}{\cosh{(r(t))}}$ and $l=\frac{1}{2}(m+n)$.

\section{Symmetries of the transition probabilities and their implications}
\label{sec:symmetries}

Let us consider the case in which, at the initial time $t=0$, we make an energy measurement on the QHO and we record the value $E_n = \hbar\omega(0)\left( n+\frac{1}{2}\right)$. From the postulate of quantum mechanics, the state of the system after the measurement is described by $\ket{n;0}_H$.
Then, the QHO evolves up to $t=\tau$, while the frequency changes from $\omega(0)$ to $\omega(\tau)$ according to a specific protocol $\omega(t)$, $t\in[0,\tau]$.
At the final time $t=\tau$, we measure the energy of the system again, obtaining the result $E_m(\tau)$. This operational sequence describes a forward process in the two-point measurement scheme (TPM), commonly adopted in quantum thermodynamics~\cite{Campisi_2011,Esposito2009,GherardiniChaos2022}. The conditional probability of obtaining $E_m(\tau)$ in the forward process, after have observed $E_n(0)$ at the initial time is
\begin{equation}
p^F_{m|n} = \left| {}_{H}\!\bra{m;\tau}\hat{U}^F(\tau,t)\ket{n;0}_H \right|^{2}
\end{equation}
with $\hat{U}^F(\tau,0)$ given by \eqref{eq:QHOEvolutionOperator}.

Now we describe the backward process. We start introducing the reverse frequency protocol, which we denote as $\tilde{\omega}(t)$ with $t\in[0,\tau]$. By definition, $\tilde{\omega}(t) = \omega(\tau-t)$, thus retracing back the values of the frequency in time-reversed order. To properly define the backward process, we need to introduce the time-reversal operation. Time-reversal is realized in quantum mechanics by the anti-unitary time-reversal operator $\hat{\Theta}$. Anti-unitary means that, for every states $\ket{\phi_1},\ket{\phi_2}$:
\begin{equation}
\hat{\Theta}\left( \alpha\ket{\phi_1} + \beta\ket{\phi}_2 \right) = \alpha^*\hat{\Theta}\ket{\phi_1} + \beta^*\hat{\Theta}\ket{\phi_2}
\end{equation}
with $\hat{\Theta}\hat{\Theta}^\dagger = \hat{\Theta}^\dagger\hat{\Theta} = \hat{\mathbb{I}}$. Under time-reversal, the position and momentum operators transform as
\begin{equation}
    \hat{\Theta}\hat{q}\hat{\Theta}=\hat{q} \quad \text{and} \quad \hat{\Theta}\hat{p}\hat{\Theta}^\dagger=-\hat{p}.
\end{equation}
Notice that the Hamiltonian of the QHO is invariant under time-reversal transformations:
\begin{equation}\label{Time-ReversalInvariance}
    \hat{\Theta}\hat{H}(t)\hat{\Theta}^\dagger = \hat{H}(t)\quad \forall \, t\in[0,\tau].
\end{equation}
Experimentally, the backward process requires that, at the initial time $t=0$, $\hat{H}(\tau)$ is measured from which $E_m(\tau)$ is obtained. Then, the QHO is made evolve under the reverse frequency protocol $\tilde{\omega}(t)$ up to $t=\tau$. At the final time $\tau$, we measure $\hat{H}(0)$ obtaining $E_n(0)$.
The associated conditional probability is now given by
\begin{equation}
    p^B_{n|m}=\left|{}_{H}\!\bra{n;0}\hat{U}^B(\tau,0)\ket{m;\tau}_H\right|^{2},
\end{equation}
where $\hat{U}^B(\tau,0)$ is the evolution operator with the reverse frequency protocol.

Using solely the property of time-reversal invariance of the Hamiltonian (\ref{Time-ReversalInvariance}), a relation between the evolution operators in the forward and backward processes can be derived \cite{Campisi_2011}:
\begin{equation}\label{microreversibilityprinciple}
    \hat{U}^B(t,0) = \hat{\Theta}\hat{U}^F(\tau-t,\tau)\hat{\Theta}^{\dagger}.
\end{equation}
This defines the \textit{micro-reversibility principle} for non-autonomous quantum systems.

Thanks to the time-reversal symmetry (\ref{Time-ReversalInvariance}), the instantaneous eigenstates of $\hat{H}(t)$ just acquire a phase under the action of $\Theta$. Hence, specifying \eqref{microreversibilityprinciple} for $t=\tau$, we can write
\begin{equation}\label{eq: TimeReversalSymmetryTransitionProbability}
    p^B_{n|m}=\left|{}_{H}\!\bra{n;0}\hat{U}^B(\tau,0)\ket{m;\tau}_H \right|^2 = \left|{}_{H}\!\bra{n;0}\hat{\Theta}\,\hat{U}^{F\,\dagger}(\tau,0)\hat{\Theta}^\dagger\ket{m;\tau}_H
    \right|^2 = \left|
    {}_{H}\!\bra{n;0}\hat{U}^{F\,\dagger}(\tau,0)\ket{m;\tau}_H
    \right|^2 = p^F_{m|n}.
\end{equation}
As a result, we have demonstrated the symmetry relation $p^F_{m|n}=p^B_{n|m}$ between the conditional probabilities of the forward and backward processes.
This symmetry holds for every driven system with a time-dependent Hamiltonian that satisfies \eqref{Time-ReversalInvariance}, and it is at the core of the celebrated Tasaki-Crooks relation~\cite{Campisi_2011,Esposito2009}.

If we consider frequency protocol which are time-symmetric, namely $\tilde{\omega}(t) = \omega(t)$, then it follows that
\begin{equation}
\hat{U}^B(t,0) = \,\hat{U}^F(t,0),
\end{equation}
such that $p^B_{n|m}=p^F_{n|m}$. This leads, together with \eqref{eq: TimeReversalSymmetryTransitionProbability}, to a symmetry of the forward probabilities with respect to exchange of $n$ and $m$: $p^F_{m|n} = p^F_{n|m}$, which is respected by any quantum system driven by a symmetric protocol, under the assumption of time-reversal symmetry of the Hamiltonian.
Interestingly, for a QHO, the relation $p^F_{m|n} = p^F_{n|m}$ remains valid also for a non-symmetric protocol. To determine this, let us consider the explicit expression of the conditional probability $p^F_{m|n}$, which is given by the squared modulus of the corresponding transition amplitude. From Sec.~\ref{Section:Exitationampl}, we have:
\begin{equation}
    p^F_{m|n}=\left| {}_{H}\!\bra{m;t}\hat{S}(t,\xi)\ket{n;t}_H\right|^2
\end{equation}
with $\hat{S}(t,\xi)$ denoting the squeezing operator.
The squared modulus of the matrix elements of $\hat{S}(t,\xi)$ depends only on the modulus of the squeezing parameter, $|r|$ [see \eqref{SqueezingParameters_ModulusPhase}]. Hence, from \eqref{SqueezingFockLegendre}, we can write:
\begin{equation}\label{eq:transitionprobLegendre}
    p^F_{m|n}=\frac{\min(m,n)!}{\max(m,n)!} x(t) \left|\mathcal{P}^{|k|}_l \big( x(t) \big)\right|^2
\end{equation}
with $x(t) = \frac{1}{\cosh(r(t))}$, $l = \frac{m+n}{2}$ and $k = \frac{m-n}{2}$. \eqref{eq:transitionprobLegendre} is invariant with respect to the exchange of the initial and final states. This symmetry was first noted in \cite{Popov1969}, and it has been exploited in \cite{Ford2012} in a quantum thermodynamics context. 

%%%%%%%%%%%%%%%%%%%%%%%%%%%%%%%%%%%%%%%%
\section{Adiabaticity breaking under squeezing production}
\label{section: AdiabaticFactor}

The theory developed in the previous sections is valid for a generic frequency protocol. The \textit{quantum adiabatic theorem}~\cite{messiah2014quantum,kato1950adiabatic} states that, when a change in $\hat{H}(t)$ is made sufficiently slow, the system, initially in a stationary state of $\hat{H}(t_0)$, evolves passing through stationary states of $\hat{H}(t)$ for all $t$. A non-adiabatic dynamic is thus associated to the creation of excitations, meaning that at a given instant the system is in a superposition of the instantaneous eigenstates of $\hat{H}(t)$.

The degree of adiabaticity of the dynamics is characterized by the \textit{adiabaticity factor}:
\begin{equation}\label{AdiabFactor}
    Q(t)=\frac{1}{2\omega(t)}\left(  M\dot{\sigma}^2(t)+M\omega^2(t)\sigma^2(t)+\frac{1}{M\sigma^2(t)}\right).
\end{equation}
In order to understand the origin and the physical meaning of the adiabaticity factor, we consider again the special case in which the QHO is in the ground state of $\hat{H}(t_0)$ at the initial time $t_0$. Then, we compute the expectation value $\bra{\psi(t)}\hat{H}(t)\ket{\psi(t)}$, where the dynamical state of the QHO $\ket{\psi(t)}$ is the dynamical state of the system that is given by \eqref{ket_solution} with $c_n=\delta_{n,0}$ (thus, the QHO is initialized in the ground state):
\begin{equation}
    \ket{\psi(t)}=e^{i\alpha_0(t)}\ket{0;t}_{I}.
\end{equation}
This entails that $\bra{\psi(t)}\hat{H}(t)\ket{\psi(t)} ={}_{I}\!\bra{0;t}\hat{H}(t)\ket{0;t}_{I}.$
Using the expression of the Hamiltonian at a generic time $t$ [\eqref{Hamiltonian_InvariantOperatorsExpression}], we get~\footnote{Note that this result was previously derived in the treatment of the phase term for a generic number states, \eqref{averageH}.}:
\begin{equation}\label{eq:averageHGS}
    \bra{\psi(t)}\hat{H}(t)\ket{\psi(t)} = \frac{\hbar}{4}\left(   M\dot{\sigma}^2(t)+M\omega^2(t)\sigma^2(t)+\frac{1}{M\sigma^2(t)}\right)
\end{equation}
For an adiabatic process we know that $\bra{\psi(t)}\hat{H}(t)\ket{\psi(t)}_{ad}\equiv\frac{\hbar\omega(t)}{2}$. Instead, for generic non-adiabatic processes, if we use the definition \eqref{AdiabFactor} for the adiabaticity factor, \eqref{eq:averageHGS} becomes
\begin{equation}\label{averageEnergyGS}
     \bra{\psi(t)}\hat{H}(t)\ket{\psi(t)}=Q(t) \frac{\hbar\omega(t)}{2}.
\end{equation}
Hence, $Q(t)$ is the ratio between the non-adiabatic and adiabatic mean energy. We remark that this property, which has been derived assuming the QHO in its ground state at $t=0$, continues to hold if $\ket{\psi(0)}$ is an eigenstate of $\hat{H}(0)$.
In the adiabatic limit, which corresponds to $\frac{\dot{\omega}(t)}{\omega^2(t)}\ll 1$~\cite{L-R1969}, whereby $\ddot{\sigma}(t)\approx 0$,
\begin{equation}\label{Ermakov_AdiabaticSolution}
    \sigma_{ad}(t)=\frac{1}{\sqrt{\omega(t)}}.
\end{equation}
Plugging \eqref{Ermakov_AdiabaticSolution} into \eqref{AdiabFactor}, we get $Q(t)=1$, as it should be in the adiabatic limit. Moreover, following the argument of Husimi~\cite{Husimi}, we can also show that $Q(t)\ge 1$ for all $t$. To do this, let's define the random variable $\nu(t)$ that denotes the number of excitations at time $t$, such that $\bra{\psi(t)}\hat{H}(t)\ket{\psi(t)}=\hbar\omega(t)\left(  \langle\nu(t)\rangle+\frac{1}{2}\right)$. From the positivity of the spectrum of the QHO's Hamiltonian, the average value $\langle\nu(t)\rangle\ge0$. Thus, from \eqref{averageEnergyGS} we get: 
\begin{equation}\label{vacuumexcitationnumber}
    \left\langle  \nu(t)\right\rangle = \frac{1}{2} \left( Q\left(t\right)-1 \right) \ge 0
\end{equation}
that implies $Q(t)\ge 1$. In conclusion, $Q(t)$ is a measure of the adiabaticity of the driving protocol: the more it deviates from $1$, the more the process can be considered non-adiabatic.

From Eqs.~(\ref{BogoliubovExplicitCoefficients}) and (\ref{eq:real_parameters}), we can relate the modulus $r(t)$ of the squeezing parameter to $Q(t)$ at any time $t$:
\begin{equation}\label{r_versus_Q}
    \begin{cases}
        \displaystyle{\cosh(2r(t)) = Q(t)} \\ 
        \displaystyle{\cosh^2(r(t)) = \frac{Q(t)+1}{2}} \\ 
        \displaystyle{\sinh^2(r(t)) = \frac{Q(t)-1}{2}.}
    \end{cases}
\end{equation}
As observed in Ref.~\cite{GalveLutz2009}, \eqref{r_versus_Q} is an important result that directly connects the degree of squeezing of the QHO's dynamics to the non-adiabaticity parameter, and thus to the frequency modulation $\omega(t)$. Specifically, \eqref{r_versus_Q} entails that the squeezing production requires a non-adiabatic change
in the frequency, i.e., $Q(t)\ge 1$.

%%%%%%%%%%%%%%%%%%%%%%%%%%%%%%%%%%%%%%%%%%%%%%%%%%%%%%%%%
\subsection{Numerical study with hyperbolic tangent functions}

%%%%%%%%%%%%%%%%%%%%%%%%%%%%%%%%%%%%%%%%%%%%%%%%%%%%%%%%%
% Figure 2 - hyperbolic quench
%%%%%%%%%%%%%%%%%%%%%%%%%%%%%%%%%%%%%%%%%%%%%%%%%%%%%%%%%
\begin{figure*}
    \centering
    \includegraphics[width=0.32\columnwidth]{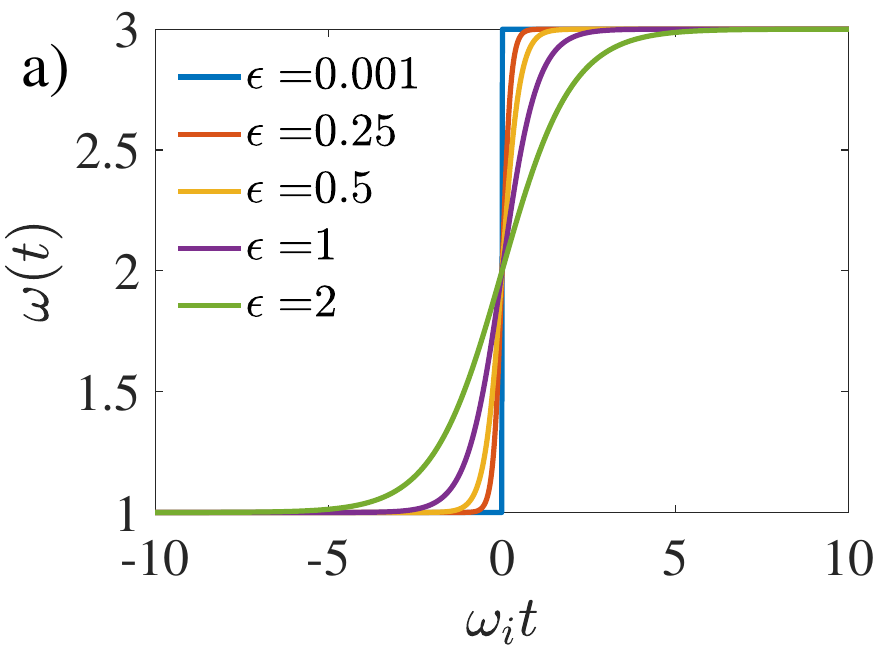}
    \includegraphics[width=0.33\columnwidth]{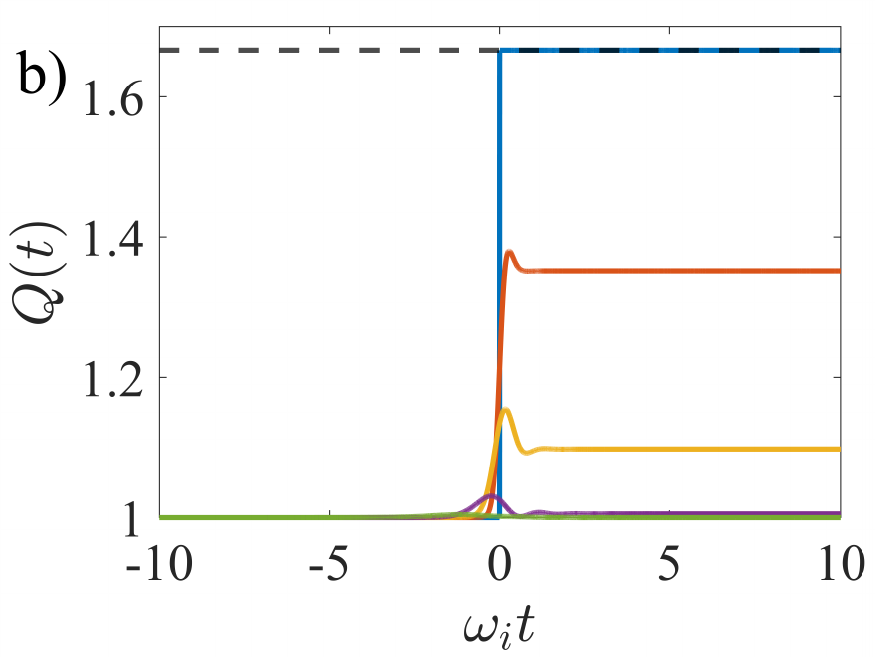}
    \includegraphics[width=0.33\columnwidth]{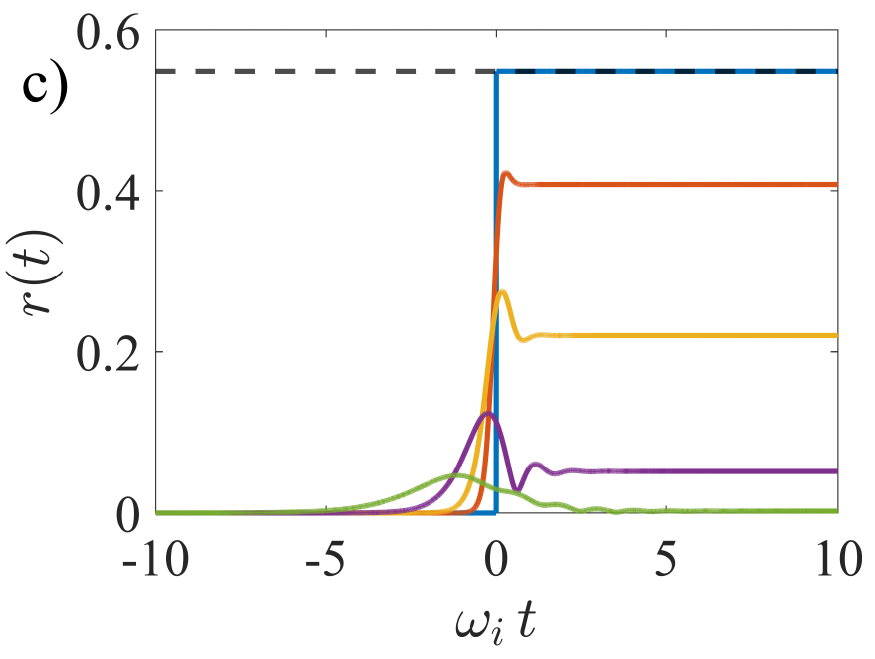}
    \caption{
    Time evolution of the QHO under an hyperbolic quench with parameters $\omega_i=1$, $\omega_f=3$, $\tau=0$ (in arbitrary units with $\hbar=M=1$). 
    (a) Frequency protocol $\omega(t)$ as a function of time for $\epsilon = 0.001, 0.25, 0.5, 1, 2$ [see \eqref{Hyperbolicprotocol}]. 
    (b) Adiabaticity factor $Q(t)$ [\eqref{AdiabFactor}] as a function of time. 
    (c) Modulus of the squeezing parameter $r(t)$ as a function of time. 
    In panels (b) and (c), the values of $\epsilon$ are the same as those used in panel (a).
    }    
    \label{fig:hyperbolicquench}
\end{figure*}
%%%%%%%%%%%%%%%%%%%%%%%%%%%%%%%%%%%%%%%%%%%%%%%%%%%%%%%%%

We model the frequency protocol using the following family of hyperbolic tangent functions~\cite{Quantum_circuit,Martinez_Tibaduiza_2021} [see panel (a) of Fig.~\ref{fig:hyperbolicquench}]:
\begin{equation}\label{Hyperbolicprotocol}
    \omega(t)=\frac{\omega_i+\omega_f}{2}+\frac{\omega_i-\omega_f}{2}\tanh\left(\frac{t-\tau}{\epsilon}\right),
\end{equation}
where $\omega_i$ and $\omega_f$ are the asymptotic frequencies reached in the limits of remote past ($t=-\infty$) and distant future ($t=+\infty)$, respectively. The parameter $\tau$ defines the instant at which the QHO's frequency assumes the mean value between $\omega_i$ and $\omega_f$, while $\epsilon$ is a continuous parameter (with dimension of time) that controls the transition rate of $\omega(t)$ [see panel (a) of Fig.~\ref{fig:hyperbolicquench}].
In fact, from \eqref{Hyperbolicprotocol}, we can write
\begin{equation}
    \epsilon=\frac{\omega_i-\omega_f}{2\dot{\omega}(\tau)},
\end{equation}
where it is evident that the limiting case $\epsilon \to 0$ corresponds to a sudden change at $t=\tau$, while $\epsilon\to \infty$ denotes a smooth transition, representing the adiabatic limit.

We obtain the time-evolution of the adiabaticity factor $Q(t)$ and the squeezing parameter $r(t)$ by numerically solving the Ermakov equation \eqref{EPequation2} (with $M=1$) for $\sigma(t)$ and then applying \eqref{AdiabFactor} and \eqref{SqueezingParameters_ModulusPhase}. We impose the boundary conditions in the limit of remote past $\sigma(t\to-\infty)=\frac{1}{\sqrt{\omega_i}}$ and $\dot{\sigma}(t\to-\infty)=0$. 
Our numerical results are reported in panels (b) and (c) of Fig.~\ref{fig:hyperbolicquench}. From Fig.~\ref{fig:hyperbolicquench}(b), we observe that for sufficiently high values of $\epsilon$, the adiabaticity factor remains $Q(t)\simeq 1$ throughout the protocol. In contrast, the sudden quench limit, characterized by $Q>1$, is achieved as $\epsilon \to 0 $ (see Sec.~\ref{Section:SuddenQuenchsection}). 
As a consequence, Fig.~\ref{fig:hyperbolicquench}(c) shows the direct correlation between the adiabaticity breaking and the generation of squeezing: the more the protocol drives the system out of equilibrium, the more the squeezing produced in the final state. 

%%%%%%%%%%%%%%%%%%%%%%%%%%%%%%%%%%%%%%%%
\section{Solutions of the Ermakov equation}
\label{sec:solution_Ermakov}

This section is dedicated to the solution of the nonlinear differential equation
\begin{equation}\label{Ermakov_Pinney}
    \ddot{\sigma}(t)+\omega^2(t)\sigma(t)=\frac{c}{\sigma^3(t)},
\end{equation}
known as the \textit{Ermakov equation}, where the constant term $c$ here is identified as $1/M^2$. While this equation was first investigated by the russian mathematician V.P.~Ermakov in 1880~\cite{ermakov1880}, we report here first the solution method provided by E.~Pinney~\cite{pinney1950}. 

%%%%%%%%%%%%%%%%%%%%%%%%%%%%%%%%%%%%%%%%
\subsection{Pinney solution}

In Ref.~\cite{pinney1950}, the solution of \eqref{Ermakov_Pinney} is expressed in terms of a fundamental set of solutions to the associated homogeneous equation, which corresponds to the equation of motion for a QHO with a time-dependent frequency:
\begin{equation}\label{HO_eq}
    \ddot{x} + \omega^2(t)x=0.
\end{equation}
For any two linearly independent solutions $x_1(t)$ and $x_2(t)$, we define their Wronskian, ${\rm Wr} [x_1,x_2]$, as the determinant:
\begin{equation}
    {\rm Wr} [x_1, x_2] = \det 
    \begin{pmatrix} 
    x_1(t) & x_2(t) \\
    \dot{x}_1(t) & \dot{x}_2(t) 
    \end{pmatrix} 
    = x_1(t)\dot{x}_2(t) - \dot{x}_1(t)x_2(t) \,.
\end{equation}
Due to Abel's Identity, for a second-order differential equation lacking a first-derivative term (like \eqref{HO_eq}), the Wronskian is a constant of motion, meaning its value is independent of time. 
By denoting this set of solutions as $\{ x_1(t),x_2(t) \}$, Pinney obtained the general solution of the Ermakov equation in the form
\begin{equation}\label{Pinney_solution}
    \sigma(t) = \left( Ax_1^2+2Bx_1x_2+Cx_2^2 \right)^{\frac{1}{2}}
\end{equation}
with the coefficients satisfying $AC - B^2 = \frac{c}{\left({\rm Wr}[x_1,x_2]\right)^2}$.

To verify that \eqref{Pinney_solution} is indeed a valid solution to the Ermakov equation, we use the identity
\begin{equation}
    \sigma\dot{\sigma} = A x_1\dot{x_1} + B(x_1\dot{x_2} + x_2\dot{x_1}) + Cx_2\dot{x_2},
\end{equation}
together with the condition that $x_1$ and $x_2$ satisfy \eqref{HO_eq}. Through some algebraic manipulation, we derive:
\begin{equation}
    \sigma \left( \ddot{\sigma} + \omega^2(t)\sigma \right) = \frac{(AC-B^2){\rm Wr}^2\left[ x_1,x_2 \right]}{\sigma^2} = \frac{c}{\sigma^2}\,,
\end{equation}
thus confirming that \eqref{Pinney_solution} satisfies the Ermakov equation.

Let us now present the solution of \eqref{Ermakov_Pinney} for a specific set of initial conditions. In particular, we consider the following Cauchy problem:
\begin{equation}\label{Ermakov_Cauchy}
    \begin{cases}
        \ddot{\sigma}(t) +\omega^2(t)\sigma(t) = \frac{c}{\sigma^3(t)} \\ 
        \sigma(0) = \sigma_0\ne0\\ 
        \dot{\sigma}(0)=\dot{\sigma}_0 \,, 
    \end{cases}
\end{equation}
where $\sigma_0$ and $\dot{\sigma}_0$ are real constants. We choose two independent solutions $x_1(t)$ and $x_2(t)$ of \eqref{HO_eq} such that
\begin{equation}
    \begin{cases}
        x_1(0)=\sigma_0 \\ 
        \dot{x}_1(0)=\dot{\sigma}_0 
    \end{cases}
    \quad \text{and} \quad
    \begin{cases}
        x_2(0)=0 \\ 
        \dot{x}_2(0) = 1/\sigma_0
    \end{cases}
\end{equation}
yielding a Wronskian of ${\rm Wr} \left[ x_1,x_2 \right] = +1$.
It can be easily verified by direct substitution that the solution to the Cauchy problem in \eqref{Ermakov_Cauchy} is given by
\begin{equation}\label{exactsolcauchyErmakov}
    \sigma(t) = \left[ x_1^2(t) + c\,x_2^2(t) \right]^{\frac{1}{2}} \,.
\end{equation}

%%%%%%%%%%%%%%%%%%%%%%%%%%%%%%%%%%%%%%%%
\subsection{Alternative construction of the Ermakov solution}

In this section, we present a construction of the exact solutions to the Ermakov equation based on complex solutions of the associated homogeneous equation \ref{HO_eq}. Following the conventions established in Refs.~\cite{Kim_2016,Defenu_2018,Defenu_2021QAC}, we set the Ermakov constant $c = \frac{1}{4}$. As a consequence, we consider the differential equation
\begin{equation}\label{ErmakovPinneyDefenu}
    \ddot{\sigma}+\omega^2(t)\sigma=\frac{1}{4\sigma^3} \,.
\end{equation}
The fundamental observation is that the Pinney solution can be generated by a complex solution $w(t)$ of \eqref{HO_eq}. We define this complex solution as
\begin{equation}\label{complex solution QHO diff eq}
    w(t)=\sqrt{A}x_1(t)+\frac{B-i\sqrt{AC-B^2}}{\sqrt{A}}x_2(t) \,,
\end{equation}
where $x_1(t)$ and $x_2(t)$ are two linearly independent real solutions of \eqref{HO_eq}. To ensure that this construction satisfies \eqref{ErmakovPinneyDefenu}, the coefficients must satisfy the following relation derived from the Pinney identity
\begin{equation}\label{PinneyWronskian}
    AC-B^2 = \left( \frac{1}{2 {\rm Wr}\left[x_1(t),x_2(t) \right]} \right)^{2} \,.
\end{equation}
By construction, the modulus $\sigma(t)=\left|w(t)\right|$ satisfies \eqref{ErmakovPinneyDefenu}. Furthermore \eqref{PinneyWronskian} imposes a normalization condition on the Wronskian of $w(t)$ and its complex conjugate
\begin{equation}\label{Wronskiancond}
   {\rm Wr}\left[w(t),w^*(t)\right] = 2i\sqrt{AC-B^2} \ {\rm Wr}\left[ x_1(t), x_2(t)\right] = -i \,.
\end{equation}

We can reverse the above logic to construct a solution $\sigma(t)$ through the following steps:
\begin{itemize}
    \item 
    Find two linearly independent real solutions, $x_1(t)$ and $x_2(t)$, of the homogeneous equation \ref{HO_eq}.
    \item 
    Construct a complex solution $w(t)=ax_1(t)+bx_2(t)$ where $a \in \mathbb{R}$ and $b \in \mathbb{C}$.
    \item 
    Impose the normalization condition on the Wronskian ${\rm Wr}\left[w(t),w^*(t)\right]=-i$.
    \item 
    The Ermakov solution is then given by the magnitude of the complex solution $\sigma(t) = \left|w(t)\right| = \sqrt{w^*(t)w(t)}$. 
\end{itemize}

%%%%%%%%%%%%%%%%%%%%%%%%%%%%%%%%%%%%%%%%
\section{Sudden and linear quench protocols}
\label{sec:protocols}

\subsection{Sudden quench}
\label{Section:SuddenQuenchsection}
As a first application of the results derived above, we find the analytical solution of \eqref{Ermakov_Pinney} for a sudden quench, i.e., an instantaneous jump in frequency. This represents the simplest frequency variation protocol: at a given instant of time, $t=0$, the frequency is abruptly changed from an initial value $\omega_i$ to a final value $\omega_f$.

\begin{figure}
    \centering
    \includegraphics[width=0.4\columnwidth]{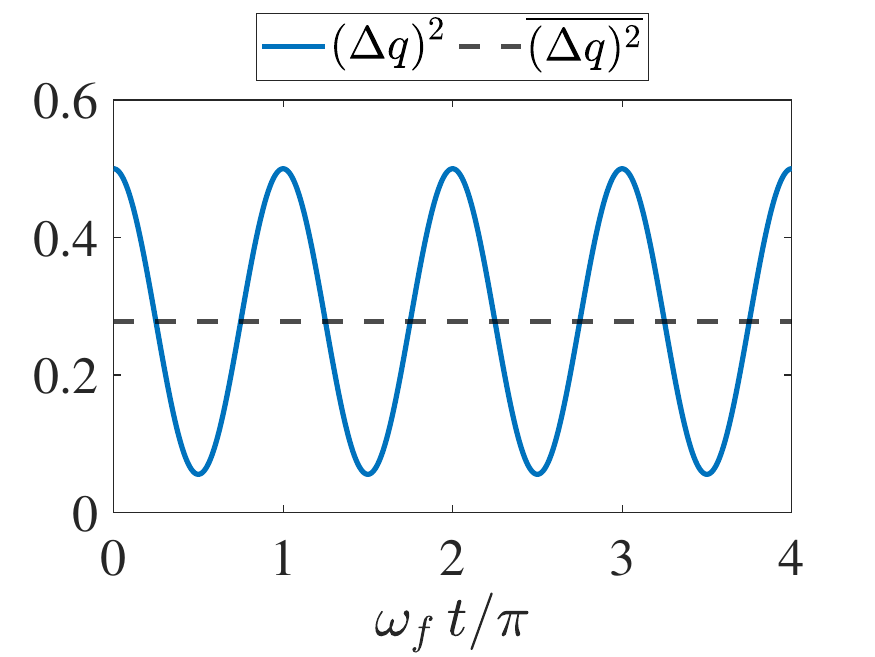}
    \includegraphics[width=0.4\columnwidth]{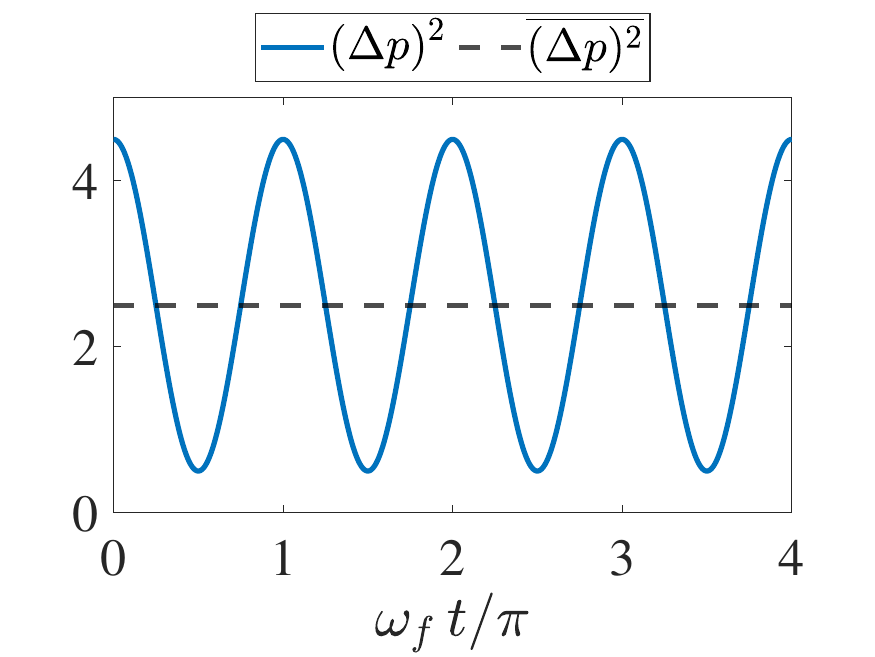}
    \caption{Position and momentum variances for a sudden quench protocol as functions of time for $n=0$ (blue solid lines), together with their values averaged over one period (black dashed line), as defined in \eqref{eq:variances_pos_mom} and \eqref{eq:average_variances_pos_mom}, respectively. 
    The pre- and post-quench frequencies are set to $\omega_i=1$ and $\omega_f=3$ (in arbitrary units with $\hbar=M=1$).}
    \label{fig:momentum_squeezing_quench}
\end{figure}

We set $c=\frac{1}{M^2}$ in \eqref{Ermakov_Pinney} and impose initial conditions for the auxiliary function $\sigma(0)=\frac{1}{\sqrt{M\omega_i}}$ and $\dot{\sigma}(0)=0$. These choices ensure that, at the initial time, the Lewis-Riesenfeld invariant and the Hamiltonian share the same eigenstates. In order to apply the Cauchy solution in \eqref{exactsolcauchyErmakov}, we have to find two solutions to the homogeneous equation
\begin{equation}
    \ddot{x}+\omega_f^2x=0
\end{equation}
subject to the following initial conditions
\begin{equation}
    \begin{cases}
        \displaystyle{x_1(0)=\frac{1}{\sqrt{M\omega_i}}} \\ 
        \displaystyle{\dot{x}_1(0)=0 } 
    \end{cases}
    \quad \text{and} \quad
    \begin{cases}    
        \displaystyle{x_2(0)=0} \\ 
        \displaystyle{\dot{x}_2(0)=\sqrt{M\omega_i} }\,. 
    \end{cases}
\end{equation}
As a result, the two independent solutions are:
\begin{equation}\label{linearindipsolconstHO}
    \begin{cases}
        \displaystyle{x_1(t)=\frac{1}{\sqrt{M\omega_i}}\cos(\omega_ft)} \\ 
        \displaystyle{{x}_2(t)=\frac{\sqrt{M\omega_i}}{\omega_f}\sin(\omega_ft) } \,. 
    \end{cases}
\end{equation}
Substituting \eqref{linearindipsolconstHO} into \eqref{exactsolcauchyErmakov} we obtain the explicit time dependence of the auxiliary variable:
\begin{equation}
        \displaystyle{\sigma(t)=\frac{1}{\sqrt{M\omega_i}}\left[\cos^2(\omega_ft)+\frac{\omega_i^2}{\omega_f^2}\sin^2(\omega_ft)\right] }^{\frac{1}{2}} \,.
\end{equation}
With the exact time dependence of $\sigma(t)$ established, we can evaluate the \textit{adiabaticity factor} and the squeezing parameters using \eqref{SqueezingParameters_ModulusPhase}:
\begin{equation}
    \begin{cases}
    Q^{\rm quench}=\frac{\omega_i^2+\omega_f^2}{2\omega_i\omega_f}\\
    r^{\rm quench}=\cosh^{-1}\left\{ \frac{\omega_i+\omega_f}{2\sqrt{\omega_i\omega_f}} \right\}\\
    \cos{\left(\phi^{\rm quench}\right)}=\rm sign(\omega_i-\omega_f)\cos(2\omega_ft).
    \end{cases}
\end{equation}
Note that both $Q^{\rm quench}$ and $r^{\rm quench}$ remain constants for $t>0$, while the squeezing phase $\phi^{\rm quench}$ changes linearly in time. As a consequence, we expect the position and momentum variances to exhibit periodic behavior (see Fig.~\ref{fig:momentum_squeezing_quench}).
In particular, from \eqref{position-momentumvariance}, we find
\begin{equation}\label{eq:variances_pos_mom}
\begin{cases}
    (\Delta q)_n^2(t)=\frac{\hbar}{M\omega_i}\left(n+\frac{1}{2}\right)\left[\cos^2(\omega_ft)+\frac{\omega_i^2}{\omega_f^2} \sin^2(\omega_ft) \right]\\ \\ 
    (\Delta p)_n^2(t)=\hbar M\omega_i\left(n+\frac{1}{2}\right)\left[\sin^2(\omega_ft)+\frac{\omega_f^2}{\omega_i^2}\cos^2(\omega_ft)\right].
\end{cases}
\end{equation}
Finally, by denoting the time average over one period as $\overline{O}$, we can write the mean variances as:
\begin{equation}\label{eq:average_variances_pos_mom}
    \begin{cases}
        \overline{\left(\Delta q\right)_n^2(t)}=\frac{\hbar}{2M\omega_i}\left(n+\frac{1}{2}\right)\left(1+\frac{\omega_i^2}{\omega_f^2}\right)\\ \\
        \overline{\left(\Delta p\right)_n^2(t)}=\frac{\hbar M\omega_i}{2}\left(n+\frac{1}{2}\right)\left(1+\frac{\omega_f^2}{\omega_i^2}\right)
    \end{cases}
\end{equation}

%%%%%%%%%%%%%%%%%%%%%%%%%%%%%%%%%%%%%%%%%%%%%%%%%%%
% Figure 3 - Oscillations of q and p variances
%%%%%%%%%%%%%%%%%%%%%%%%%%%%%%%%%%%%%%%%%%%%%%%%%%%
%%%%%%%%%%%%%%%%%%%%%%%%%%%%%%%%%%%%%%%%%%%%%%%%%%%

%%%%%%%%%%%%%%%%%%%%%%%%%%%%%%%%%%%%%%%%
\subsection{Linear symmetric quench}

We now consider a linear frequency protocol with a quench rate $\delta>0$, defined as:
\begin{equation}\label{cyclicramp}
    \omega^2(t)=\delta\left|t\right| \,.
\end{equation}
To simplify the analysis, it is convenient to introduce the following rescaled variables 
\begin{equation}
\sigma = \delta^{-\frac{1}{6}}\tilde{\sigma} \quad \text{and} \quad t = \delta^{-\frac{1}{3}} s \,.
\end{equation}
Under this rescaling and setting $c=\frac{1}{4}$, the Ermakov equation becomes
\begin{equation}
    \ddot{\tilde{\sigma}}(s) + \tilde{\omega}^2(s) \tilde{\sigma}(s) = \frac{1}{4\tilde{\sigma}^3(s)}\,,
\end{equation}
where the rescaled frequency is $\tilde{\omega}^2(s)=\left|s\right|$. Note that physically relevant quantities, such as the modulus of the squeezing parameter $r(t)$ and the adiabaticity factor $Q(t)$ are invariant under this rescaling. 
As a consequence, the dynamics for any arbitrary rate $\delta$ can be reduced to the case $\delta=1$. In the following, we suppress the tilde over the rescaled quantities for clarity.

The classical equation of motion associated with this protocol is
\begin{equation}
    \ddot{\sigma}(s) + |s| \,\sigma(s) = 0 \,.
\end{equation}
This is a standard differential equation of mathematical physics: the Airy equation \cite{abramowitz1972handbook}, which admits two linearly independent solutions
\begin{equation}
    x_1(s) = {\rm Ai}\left(-|s|\right), \qquad x_2(s) = {\rm Bi}\left(-|s|\right),
\end{equation}
where Ai and Bi are the Airy functions. The two solutions have a known constant Wronskian: ${\rm Wr}\left[x_1(s),x_2(s)\right]=\frac{1}{\pi} $. As explained in the previous section, to solve the Ermakov equation, we construct a complex solution $w(s)$ of $\ddot{w}(s) + |s| \,w(s) = 0$. We express it as a linear combination of the two Airy solutions
\begin{equation}
    w(s) = a x_1(s) + b x_2(s),
\end{equation}
where $a\in \mathbb{R}$ and $b \in \mathbb{C}$ (specifically $b=b_1+ib_2$). To ensure that the magnitude $\sigma(s) = |w(s)|$ satisfies the Ermakov equation with $c=1/4$, we must satisfy the Wronskian condition ${\rm Wr}[w,w^*]=-i$. Using the Wronskian of the Airy functions, this condition simplifies to a relation between the real coefficients
\begin{equation}\label{firstrelationcoefficient}
    a \, b_2=\frac{\pi}{2}.
\end{equation}

Substituting $w(s) = (a x_1+b_1x_1) + i(b_2x_2)$ into $\sigma(s) = \sqrt{w \,w^*}$, yields the general Ermakov solution of \eqref{ErmakovPinneyDefenu}
\begin{equation}\label{solutionErmakovAirygeneral}
    \sigma(s) = \left( \Big( ax_1(s)+b_1x_2(s)\Big)^2 + b^2_2x_2^2(s)\right)^{\frac{1}{2}}.
\end{equation}
We assume the protocol begins at $t=-\infty$, which means imposing the boundary condition
\begin{equation}
    \lim_{s \to -\infty} \frac{1}{2\sigma^2(s)} = \omega(s) \,, \qquad \lim_{s \to -\infty} \dot{\sigma}(s) = 0 \,.
\end{equation}
Using the asymptotic expansion for the Airy functions as $s \to \infty$ \cite{abramowitz1972handbook}
\begin{equation}\label{second condition}
      x_1(s) \sim \frac{\cos\left(\frac{2}{3}|s|^{3/2} - \frac{\pi}{4}\right)}{\sqrt{\pi}|s|^{1/4}} \quad \text{and} \quad x_2(s) \sim \frac{\sin\left(\frac{2}{3}|s|^{3/2} - \frac{\pi}{4}\right)}{\sqrt{\pi}|s|^{1/4}},
\end{equation}
we find that the coefficients in \eqref{solutionErmakovAirygeneral} are fully determined as
\begin{equation}
    a = \sqrt{\frac{\pi}{2}}, \qquad b_1 = 0 , \qquad b_2 = \sqrt{\frac{\pi}{2}}.
\end{equation}
Thus, the solution for $t \in (-\infty,0]$ is:
\begin{equation}\label{solutionfornegativetimes}
    \sigma^2(s) = \frac{\pi}{2} \Big( {\rm Ai}^2 \left(-|s|\right) + {\rm Bi}^2 \left(-|s|\right) \Big).
\end{equation}
This half-ramp protocol, which ends at the critical point $\omega(0)=0$, serves as a paradigm in condensed matter theory for studying quantum dynamics and defects formation near a quantum critical point (QCP)~\cite{Dziarmaga01112010,Polkovnikov2011}. Within this framework, our results provide a natural language to discuss the Kibble-Zurek mechanism (KZM)~\cite{del_Campo_2014}, which describes the non-equilibrium dynamics of a system driven through a continuous phase transition.
Specifically, crossing a QCP with a finite rate $\delta$ results in the generation of excitations (or topological defects), whose density follows a universal power-law scaling $\delta^\theta$. 
The scaling exponent $\theta$ is a universal quantity, depending solely by the system's dynamic critical exponent $z$ and its correlation length exponent $\nu$ through the relation
\begin{equation}\label{KZscaling}
    \theta = \frac{z\nu}{1+z\nu}.
\end{equation}

The QHO with the time-dependent frequency \eqref{cyclicramp} can be regarded as an effective description of a quantum many-body system driven cyclically across its QCP, with scaling exponent $z\nu=\frac{1}{2}$. Assuming the system is initialized in its ground state, the mean energy at the end of the half-ramp protocol ($t=0$) is given by \eqref{averageEnergyGS}, which leads to
\begin{equation}\label{ExcessEnergy}
    \bra{\psi(0)}\hat{H}(0)\ket{\psi(0)} = Q(0) \frac{\hbar\omega(0)}{2} \,,
\end{equation}
where $Q(0)$ is the adiabaticity factor evaluated at $t=0$. 
In condensed matter literature, this quantity is often called the \textit{heat} or \textit{excess energy}~\cite{Polkovnikov2011} and in quantum thermodynamics, it represents the average work done on the system~\cite{Buffoni_Gherardini}. Using the definition of $Q(t)$ from \eqref{AdiabFactor} and applying the rescaling relations, the energy at $t=0$ becomes
\begin{equation}
    \bra{\psi(0)}\hat{H}(0)\ket{\psi(0)} = \frac{\hbar\delta^{1/3}}{2} \left(  \dot{\sigma}^2(0) + \omega^2(0)\sigma^2(0) + \frac{1}{4\sigma^2(0)} \right) = \frac{\hbar\delta^{1/3}}{2\sigma^2(0)} \left( \frac{1}{4} + \Big( \dot{\sigma}(0) \sigma(0) \Big)^2 \right),
\end{equation}
where we have used $\omega(0)=0$. By evaluating the Airy functions and their derivatives at the origin
\begin{equation}
    {\rm Ai}(0) = \frac{1}{3^{2/3}\Gamma(2/3)} \,, \quad
    {\rm Bi}(0) = \frac{1}{3^{1/6}\Gamma(2/3)} \,, \quad 
    {\rm Ai'}(0) = -\frac{1}{3^{1/3}\Gamma(1/3)}\,, \quad 
    {\rm Bi'}(0) = \frac{3^{1/6}}{\Gamma(1/3)},
\end{equation}
we find:
\begin{equation}
    \sigma^2(0) = \frac{2\pi}{3^{4/3} \Gamma^2(2/3)}\,, \quad \sigma(0)\dot{\sigma}(0) = \frac{\pi}{3\,\Gamma(1/3)\,\Gamma(2/3)}.
\end{equation}
Applying the Euler reflection formula $\Gamma(x)\Gamma(1-x)=\frac{\pi}{\sin(\pi x)}$, we obtain the final expression for the mean energy
\begin{equation}\label{ExcessEnergyHalfRamp}
    \bra{\psi(0)}\hat{H}(0)\ket{\psi(0)} = \frac{\pi\,\hbar}{3^{2/3}\Gamma^2(\frac{1}{3})}\delta^{1/3} \,.
\end{equation}
This result confirms the KZM scaling with $z\nu=1/2$. Such mean-field exponents are typical of quantum spin models with long-range interactions~\cite{Buffoni_Gherardini}, demonstrating that the time-dependent QHO effectively describes many-body systems driven across criticality.

%%%%%%%%%%%%%%%%%%%%%%%%%%%%%%%%%%%%%%%%
% Figure 4 - Non linear quench
%%%%%%%%%%%%%%%%%%%%%%%%%%%%%%%%%%%%%%%%
\begin{figure}
    \centering
    \includegraphics[width=0.325\linewidth]{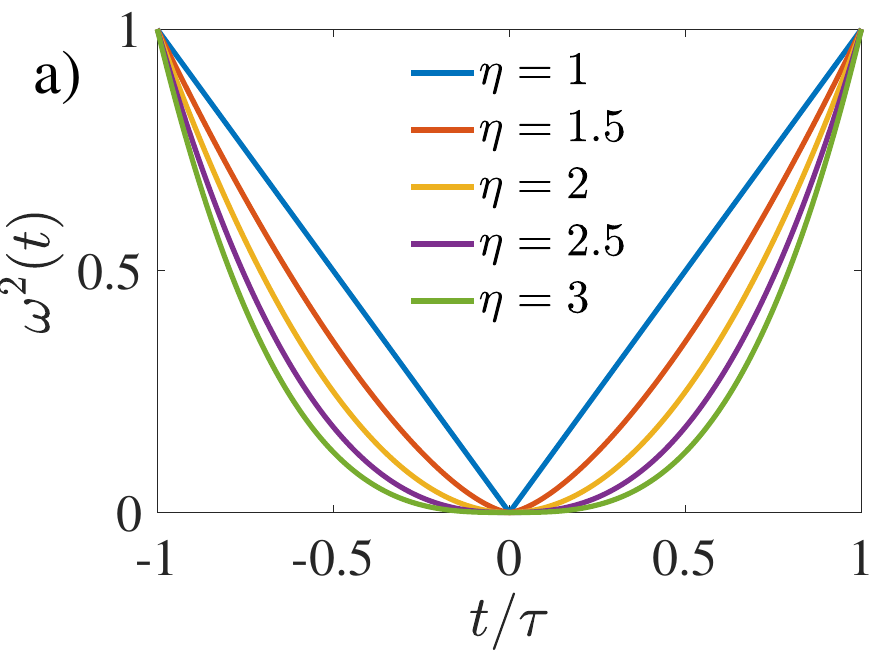}
    \includegraphics[width=0.325\linewidth]{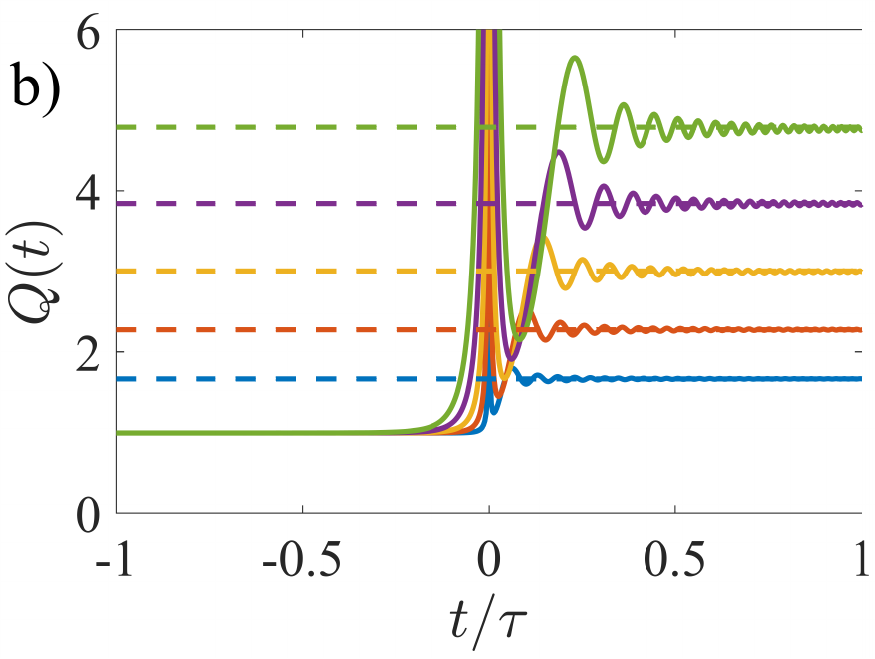}
    \includegraphics[width=0.325\linewidth]{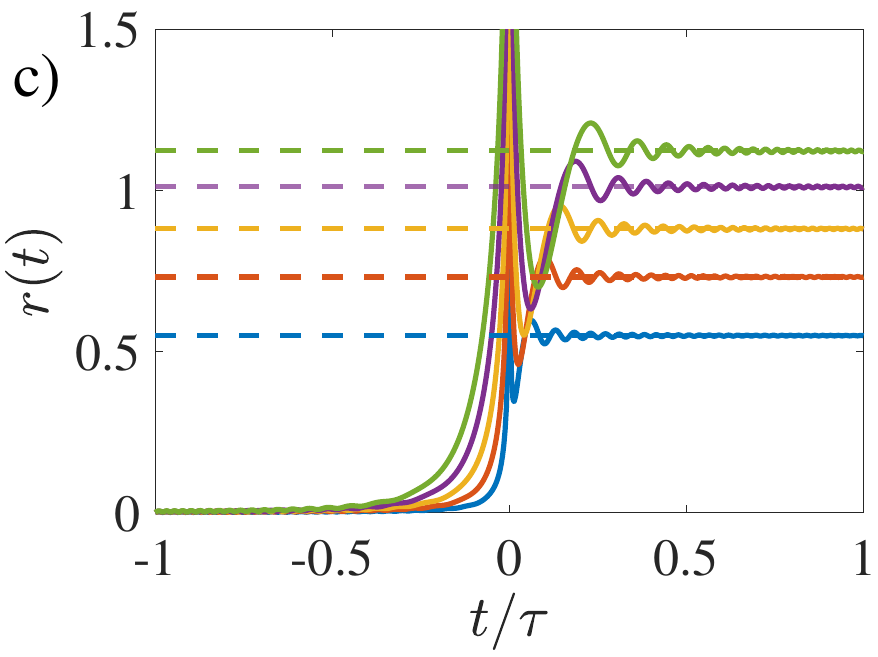}
    \caption{
    Time evolution of (a) the squared frequency protocol $\omega^2(t)$, (b) the adiabaticity factor $Q(t)$, and (c) the squeezing parameter $r(t)$ for the non-linear quench defined in \eqref{eq:nonlinear_ramp}. Different colors correspond to different values of $\eta=\{ 1,1.5,2,2.5,3 \}$, as indicated in the legend. The asymptotic values of $r(t)$ and $Q(t)$, predicted by \eqref{eq:steady_r_nonlinear}, are shown by the dashed lines. The breakdown of adiabaticity is signaled by the finite asymptotic values attained by $r(t)$ and $Q(t)-1$  at long times, in agreement with the prediction of \eqref{eq:steady_r_nonlinear}. 
    }
    \label{fig:nonlinear_quench}
\end{figure}

To obtain an exact solution for the full cyclical ramp, we extend the solution \eqref{solutionfornegativetimes} to the interval $t\in\left(0,+\infty\right)$. We only outline the procedure here, referring to Ref.~\cite{Defenu_2021QAC} for the explicit calculations.

Let us consider again a general solution of the form \eqref{General solution} and impose boundary conditions to ensure the match with the previous solution in the $t\leq 0$ interval. In this way, the Wronskian condition \eqref{firstrelationcoefficient} fully determines the coefficients. The final result for the auxiliary function in the interval $t > 0$ is
\begin{equation}\label{eq:solution_positivetime}
    \sigma^2(s) = a_+^2 {\rm Ai}^2 \left(-|s|\right) + b_+^2 {\rm Bi}^2 \left(-|s|\right),
\end{equation}
where $a_+ = \sqrt{\frac{3\pi}{2}}$ and $b_+=\sqrt{\frac{\pi}{6}}$.

Assuming that the protocol ends at $t=+\infty$, we perform the rescaling $\sigma(t)\to \sqrt{\frac{M}{2}}\sigma(t)$ in the above solution of the auxiliary function and substitute \eqref{eq:solution_positivetime} into \eqref{SqueezingParameters_ModulusPhase}. We thus obtain the asymptotic value of the squeezing parameter for the linear quench~\cite{Defenu_2021QAC}:
\begin{equation}\label{eq:steady_r_linear}
    \lim_{t \to +\infty}r(t)=\cosh^{-1}\left(\frac{2}{\sqrt{3}}\right) \approx 0.658 \,,
\end{equation}
which is visible in Fig.~\ref{fig:nonlinear_quench}(c) for the special case $\eta =1$ (blue dashed line).

This analysis can be generalized to a non-linear protocol of the form
\begin{equation}\label{eq:nonlinear_ramp}
    \omega^2(t) = \Big( \delta\left|t\right| \Big)^{\eta} \,,
\end{equation}
which reduces to the linear case for $\eta=1$ [see Fig.~\ref{fig:nonlinear_quench}(a)]. 
For this generalized protocol, we perform a numerical analysis by directly integrating the Ermakov equation \eqref{EPequation2}, using standard Runge-Kutta methods. The resulting solution $\sigma(t)$ is then employed to evaluate the adiabaticity factor $Q(t)$, defined in \eqref{AdiabFactor}, and the squeezing parameter $r(t)$, defined in \eqref{r_versus_Q} [see Fig.~\ref{fig:nonlinear_quench}(b)-(c)]. 

\begin{figure}
    \centering
    \includegraphics[width=0.45\columnwidth]{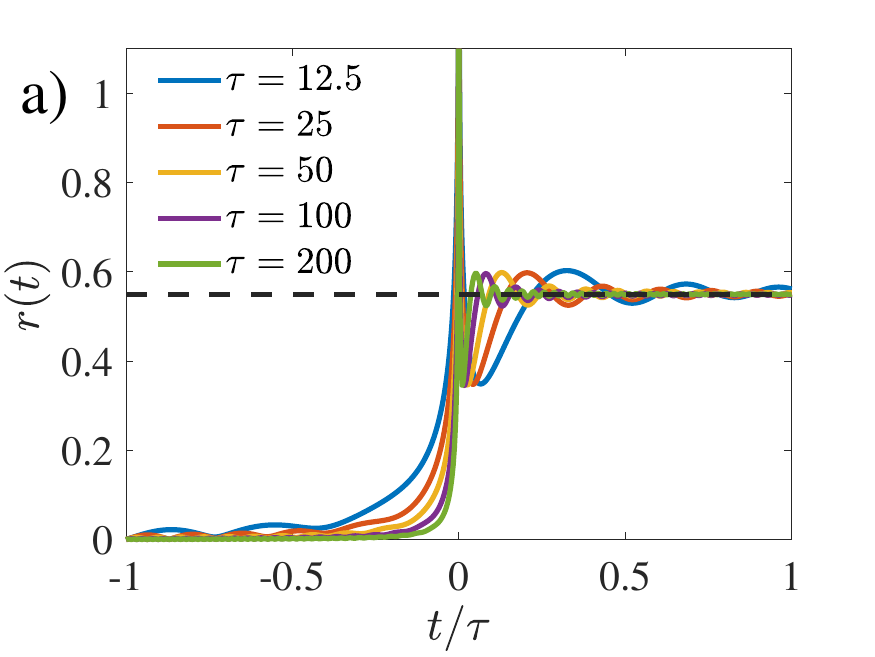}
    \includegraphics[width=0.45\columnwidth]{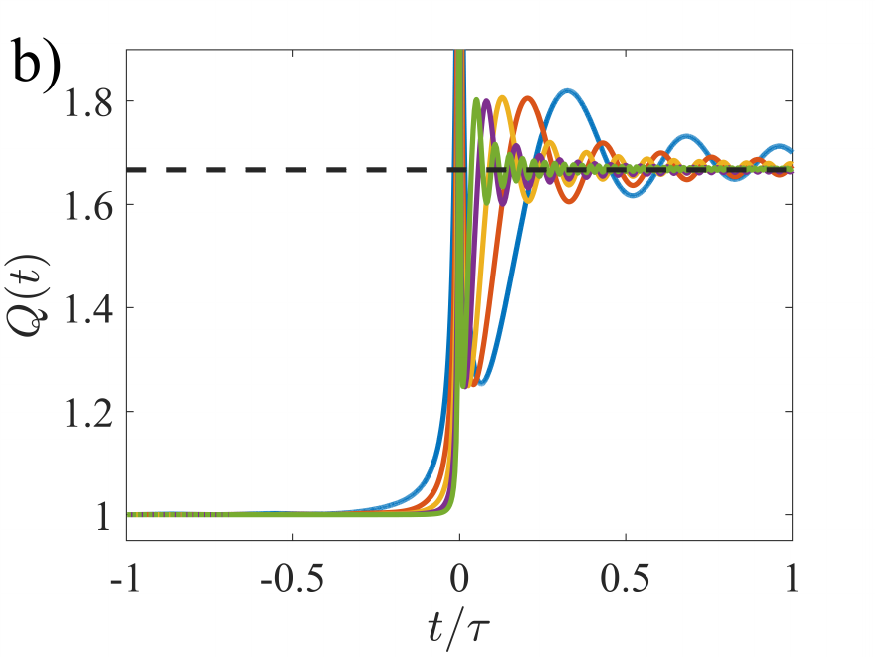}
    \caption{
    Time evolution of (a) the squeezing parameter $r(t)$ and (b) the adiabaticity factor $Q(t)$  as a function of the rescaled time $t/\tau$ for a linear quench protocol. Different colors correspond to different values of $\tau=\{12.5,25,50,100,200 \}$, as indicated in the legend (in arbitrary units with $\hbar=M=1$). 
    The black dashed line represents the theoretical prediction of \eqref{eq:steady_r_linear}. Independent of the protocol duration, the squeezing parameter $r(t)$ saturates to the non-zero value predicted by \eqref{eq:steady_r_linear}. The amplitude of the oscillations of $r(t)$, obtained from the numerical integration of \eqref{EPequation2}, decreases for increasing quench duration. In the limit $\tau\to \infty$ (or equivalently $\delta\to 0$), the numerical curves approach the analytical prediction \eqref{eq:steady_r_linear}. 
    }
    \label{fig:squeezing_linear_quench_varioustau}
\end{figure}

The asymptotic value of the squeezing parameter is predicted to be~\cite{Defenu_2021QAC}
\begin{equation}\label{eq:steady_r_nonlinear}
    \lim_{t \to +\infty}r(t) = \cosh^{-1}\left(\frac{1}{\sin{(p\pi)}}\right) \,,
\end{equation}
where $p=\frac{1}{2+\eta}$. Since $Q(t)$ is directly related to the squeezing parameter through \eqref{r_versus_Q}, the finite asymptotic value of $r(t)$ implies that $Q(t)$ also saturates to a constant value. This asymptotic value of $Q(t)$ depends solely on $p=\frac{1}{2+\eta}$ and signals the breakdown of adiabaticity whenever it exceeds unity. As shown in Fig.~\ref{fig:nonlinear_quench}, the numerical integrations of the Ermakov equation \eqref{EPequation2} for the non-linear quench protocol \eqref{eq:nonlinear_ramp} is in excellent agreement with the theoretical prediction of \eqref{eq:steady_r_nonlinear}. 
Notably, the asymptotic prediction of \eqref{eq:steady_r_nonlinear} is independent of the quench rate $\delta$. This demonstrates that the breakdown of adiabaticity persists even in the limit of an infinitely slow quench, $\delta \to 0$, as further confirmed by the numerical results for the linear protocol reported in Fig.~\ref{fig:squeezing_linear_quench_varioustau}. 

In particular, Fig.~\ref{fig:squeezing_linear_quench_varioustau} shows the time evolution of (a) the squeezing parameter $r(t)$ and (b) the adiabaticity factor $Q(t)$ as functions of the rescaled time $t/\tau$, for various quench durations $\tau = \{12.5,25,50,100,200\}$. The dynamics are obtained from a direct numerical integration of the Ermakov equation \eqref{EPequation2} for a linear ramp, with all quantities expressed in natural units ($\hbar=M=1$). The results clearly show that, despite the increasing duration of the protocol, the squeezing parameter $r(t)$ exhibits damped oscillations around a well-defined asymptotic value. This value is independent of $\tau$ and coincides with the analytical prediction given in \eqref{eq:steady_r_linear}, as highlighted by the black dashed line. The convergence toward this prediction becomes progressively more evident as $\tau$ increases, with a systematic reduction in the amplitude of the oscillations. A similar behavior is observed for the adiabaticity factor $Q(t)$, which displays transient oscillations whose magnitude decreases for slower quenches, while approaching a finite asymptotic value at long times. Importantly, this asymptotic value remains strictly larger than unity, thereby providing a direct dynamical signature of the violation of adiabaticity.

Overall, Fig.~\ref{fig:squeezing_linear_quench_varioustau} confirms that, although slower quenches reduce transient excitations, the system does not recover adiabatic behavior even in the limit $\tau \to \infty$ (equivalently $\delta \to 0$), in agreement with the analytical result of \eqref{eq:steady_r_linear}. 

%%%%%%%%%%%%%%%%%%%%%%%%%%%%%%%%%%%%%%%%
% Figure 5 
%%%%%%%%%%%%%%%%%%%%%%%%%%%%%%%%%%%%%%%%

%%%%%%%%%%%%%%%%%%%%%%%%%%%%%%%%%%%%%%%%
\section{Conclusions}

In this review paper, we have bridged together different methods and physical quantities that pertain to driven quadratic quantum systems. The common thread of this work is the connection between non-adiabaticity, squeezing production and invariants, leading to an organic summary and elaboration of results scattered across the literature. In particular, the results and derivations here reported shed light into different fields, ranging from pure mathematical physics~\cite{L-R1969,Lewis-Leach1982} to active topics like shortcuts-to-adiabaticity, quantum thermodynamics and the crossing of a quantum critical point.

Regarding shortcuts-to-adiabaticity, the derivations in the review could serve as an introduction to \textit{invariant-based inverse engineering}~\cite{GuryOdelin2019}, offering the analytical tools necessary to design fast, high-fidelity protocols in driven quantum systems.

The QHO with time-dependent frequency is also used as a \textit{working fluid} in many models of quantum heat engines~\cite{Abah2012SingleIonHeatEngine,Abah2018STAquantumEngine}. In this regard, the relation between the adiabaticity parameter $Q$ and squeezing allows for a deeper physical interpretations of the efficiency of such devices in finite-time cycles. For example, the results of Secs.~\ref{Section:Exitationampl}  
could represent a starting point to characterize the efficiency of quantum heat engine from a non-equilibrium perspective following Ref.~\cite{EfficiencyFluctuationsLutz}, but resorting to quasiprobability distributions~\cite{Gherardini_2024}.

Finally, as highlighted in Secs.~\ref{sec:protocols}, the frequency-dependent QHO can be regarded as an effective model for quantum many-body systems driven across a quantum critical point. Systems admitting such effective description are ubiquitous in condensed-matter and statistical mechanics. Examples are long-range interacting quantum spin systems~\cite{DefenuRMP2023} and in particular the Lipkin-Meshkov-Glick model~\cite{Defenu_2018}, models of interacting bosons like the bosonic Josephson junction~\cite{Gati_2007,LenaPRA2016} and models presenting superradiant quantum phase transitions as the Dike model~\cite{PueblaQuantumRabiModel,PueblaPlenioSuperradiantQPT}.

%%%%%%%%%%%%%%%%%%%%%%%%%%%%%%%%%%%%%%%%
\section*{Acknowledgments}

This work has been financial supported by the PNRR MUR project PE0000023 NQSTI funded by the European Union---Next Generation EU.

%%%%%%%%%%%%%%%%%%%%%%%%%%%%%%%%%%%%%%%%
\appendix

%%%%%%%%%%%%%%%%%%%%%%%%%%%%%%%%%%%%%%%%
\section{Diagonalization of a quadratic Hamiltonian}\label{App:SqueezingBogoliubov}

In this section, we demonstrate how a quadratic bosonic Hamiltonian can be diagonalized by means of Bogoliubov transformations, and how this procedure is equivalent to a squeezing operation. As we are assuming here a fixed time instant, we will not make explicit any time-dependence. The problem of a generic quadratic Hamiltonian can be found, for both the bosonic and fermionic case, in Refs.~\cite{ring2004nuclear,blaizot1986quantum}. 

The Hamiltonian we consider is:
\begin{equation}\label{eq:app_quadratic_Hamiltonian}
    \hat H = \eta\hat{a}^{\dagger}\hat{a} + \epsilon \hat{a}^{\dagger}\hat{a}^{\dagger}+\epsilon^{\ast} \hat{a} \hat{a}.
\end{equation}
Then, we employ the Bogoliubov transformations: $\hat{b} = u\hat{a}\ +v\hat{a}^\dagger$ and $\hat{b}^\dagger = u^*\hat{a}^\dagger+v^*\hat{a}$. In order to preserve the bosonic commutation relations, we must impose
\begin{equation}\label{bogoliubovcondition}
    \left| u \right|^2-\left| v \right|^2=1.
\end{equation}
Without loss of generality, we can take $u$ to be real and ensures the validity of \eqref{bogoliubovcondition} with the ansatz
\begin{equation}
    u = \cosh(r) \quad \text{and} \quad v=e^{i\varphi}\sinh(r),
\end{equation}
where $r$ is a positive real constant. Introducing the squeezing operator~\cite{barnett1997methods}
\begin{equation}
    \hat{S}(z)=\exp\left( -\frac{z}{2}\hat{a}^{\dagger2} +\frac{z^*}{2}\hat{a}^{2}\right),
\end{equation}
where $z=e^{i\varphi}r$, we can determine:
\begin{equation}
\begin{cases}\label{Bogoliubov_Squeezing_equiv}
    \hat b = \hat{S}(z)\hat{a}\hat{S}^{\dagger}(z)=\cosh(r)\hat{a}+e^{i\varphi}\sinh(r)\hat{a}^\dagger \\
   \hat{b}^{\dagger} = \hat{S}(z)\hat{a}^{\dagger}\hat{S}^{\dagger}(z) =\cosh(r)\hat{a}^{\dagger}+e^{-i\varphi}\sinh(r)\hat{a},
\end{cases}
\end{equation}
which outlines the equivalence between the Bogoliubov and squeezing transformations. The inverse relations are:
\begin{equation}\label{eq:inverseBT}
    \begin{cases}
         \hat{a} = \cosh(r)\hat{b} -e^{i\varphi}\sinh(r)b^{\dagger} \\
    \hat{a}^{\dagger} =\cosh(r)\hat{b}^\dagger-e^{-i\varphi}\sinh(r) \hat{b}.
    \end{cases}
\end{equation}
Inserting \eqref{eq:inverseBT} into the Hamiltonian (\ref{eq:app_quadratic_Hamiltonian}), after some algebra we determine that
\begin{eqnarray}
\hat{H} &=& \eta \left( \cosh^2(r) \hat{b}^\dagger \hat{b} + \sinh^2(r) \hat{b}\hat{b}^\dagger\right) - \frac{\eta}{2}  \sinh(2r) \left( e^{i\varphi}\hat{b}^\dagger \hat{b}^\dagger +e^{-i\varphi}\hat{b}\hat{b} \right) 
\\
&+& |\epsilon| e^{i\theta} \left( \cosh^2(r) \hat{b}^\dagger \hat{b}^\dagger + e^{-2i\varphi} \sinh^2(r) \hat{b}\hat{b} \right) - \frac{|\epsilon|}{2} e^{i(\theta-\varphi)}  \sinh(2r) (\hat{b}^\dagger \hat{b} + \hat{b} \hat{b}^\dagger)
\\
&+& |\epsilon| e^{-i\theta} \left( \cosh^2(r) \hat{b} \hat{b}+e^{2i\varphi} \sinh^2(r) \hat{b}^\dagger\hat{b}^\dagger \right) - \frac{|\epsilon|}{2} e^{-i(\theta-\varphi)}  \sinh(2r) (\hat{b}^\dagger \hat{b} + \hat{b} \hat{b}^\dagger),
\end{eqnarray}
where we have used the Euler representation of $\epsilon$ ($\epsilon=e^{i\theta}|\epsilon|$). Choosing $\varphi=\theta$, we obtain:
\begin{eqnarray}
\hat{H} &=& \Big( \eta\cosh(2r)-2|\epsilon|\sinh(2r)\Big) \hat{b}^\dagger\hat{b}\ +\left( |\epsilon|\cosh(2r)-\frac{\eta}{2}\sinh(2r)\right) \left(e^{i\theta}\hat{b}^\dagger\hat{b}^\dagger+e^{-i\theta}\hat{b}\hat{b}\right) \\
&+& \eta\sinh^2(r)-|\epsilon|\sinh(2r).
\end{eqnarray}
The non-diagonal part of $\hat{H}$ in the basis spanned by the eigenstates of $\hat{b}^\dagger\hat{b}$, can be eliminated by setting $\tanh(2r)=\frac{2|\epsilon|}{\eta}$, such that, using standard relations between hyperbolic functions, one gets:
\begin{equation}
    \hat{H}=\omega\left(\hat{b}^\dagger\hat{b}+\frac{1}{2}\right)-\frac{\eta}{2}
\end{equation}
with $\omega=\sqrt{\eta^2-4|\epsilon|^2}$, $u=\cosh(r)=\sqrt{\frac{1}{2}\left(\frac{\eta}{\omega}+1\right)}$ and $|v|=\sinh(r)=\sqrt{\frac{1}{2}\left(\frac{\eta}{\omega}-1\right)}$.
In conclusion, the ground state of $\hat{H}$ becomes $\ket{0}_H=S(z)\ket{0}$ with $\hat{a}\ket{0}=0$.

%%%%%%%%%%%%%%%%%%%%%%%%%%%%%%%%%%%%%%%%
\section{Derivation of \eqref{definitions_L}}
\label{app:definitions_L}

Let us start from the definitions of the lowering and raising operators of the invariant $\hat{I}(t)$, i.e.,\\ 
$\hat{a}(t) = \frac{1}{\sqrt{2\hbar}}\left( \frac{\hat{q}}{\sigma(t)} + i\big( \sigma(t)\hat{p} - M\dot{\sigma}(t)\hat{q} \big)\right)$ and $\hat{a}^{\dagger}(t) = \frac{1}{\sqrt{2\hbar}}\left( \frac{\hat{q}}{\sigma(t)} - i\big( \sigma(t)\hat{p} - M\dot{\sigma}(t)\hat{q}\big) \right)$.
Solving for $\hat{q}$ and $\hat{p}$ we obtain:
\begin{equation}\label{positionmomentumoperator}
\begin{cases}
    \hat{q} = \sqrt{\frac{\hbar}{2}}\sigma(t)\left(  \hat{a}(t)+\hat{a}(t)^\dagger\right)\\
    \hat{p}=\sqrt{\frac{\hbar}{2}}\left( \left( M\dot{\sigma }(t)-\frac{i}{\sigma(t)}\right)\hat{a}(t)+\left( M\dot{\sigma}(t)+\frac{i}{\sigma(t)}\right)\hat{a}^\dagger(t) \right).
\end{cases}
\end{equation}
Through some algebra, we find that
\begin{eqnarray}
\begin{cases}\label{squaredpositionmomentum}
\hat{q}^2 = \frac{\hbar}{2}\sigma^2(t) \left( \hat{a}^2(t)+\hat{a}^{\dagger2}(t)+\left\{ \hat{a}(t),\hat{a}^\dagger(t)\right\} \right)\\
\hat{p}^2 = \frac{\hbar}{2}\left( \left( M^2\dot{\sigma}^2(t)-\frac{1}{\sigma^2(t)}-2iM\frac{\dot{\sigma}(t)}{\sigma(t)}\right)\hat{a}^2(t) + \left( M^2\dot{\sigma}^2(t)-\frac{1}{\sigma^2(t)}-2iM\frac{\dot{\sigma}(t)}{\sigma(t)}\right)\hat{a}^{\dagger2}(t)
+ \left( \dot{\sigma}^2(t)+\frac{1}{\sigma^2(t)}\right)\left\{  {\hat{a}(t),\hat{a}(t)^\dagger}\right\} \right).
\end{cases}
\end{eqnarray}
Using \eqref{squaredpositionmomentum} and the Ermakov equation $\ddot{\sigma}(t)+\omega^2(t)\sigma(t)=\frac{1}{M^2\sigma^3(t)}$, \eqref{definitions_L} is derived.

Since the calculations are analogous for all the three terms, here we focus on the term proportional to $\hat{a}^2(t)$.
We recall that $\hat{H}(t) = \frac{\hat{p}^2}{2M}+\frac{1}{2}M\omega^2(t)\hat{q}^2$.
From \eqref{squaredpositionmomentum} we get, factorizing out and simplifying from the definition the constant term $\frac{\hbar}{2}$, we get that
\begin{equation}\label{eq:Lminusintermediate}
   \hat{L}_{-}(t)=\frac{1}{2}\left( M\dot{\sigma}^2(t)-\frac{1}{M\sigma^2(t)}-2i\frac{\dot{\sigma}(t)}{\sigma(t)}+M\omega^2(t)\sigma^2(t)\right)\hat{a}^2(t).
\end{equation}
Using the Ermakov equation \eqref{EPequation2}, we express 
$\frac{1}{M\sigma^2(t)}$ as $\frac{1}{M\sigma^2(t)} = M\sigma(t)\ddot{\sigma}(t)+\omega^2(t)\sigma^2(t)$. In this way, substituting into \eqref{eq:Lminusintermediate}, we finally arrive at the desired result
\begin{equation}
    \hat{L}_{-}(t) =\frac{1}{2}\left( M(\dot{\sigma}^2(t)-\sigma(t)\ddot{\sigma}(t))-2i\frac{\dot{\sigma}(t)}{\sigma(t)} \right)\hat{a}^2(t).
\end{equation}

%%%%%%%%%%%%%%%%%%%%%%%%%%%%%%%%%%%%%%%%
\section{Derivation of \eqref{adaggerpunto}}
\label{app:derivative_of_a}

Let us consider the definition of $\hat{a}^\dagger(t)$ and its time derivative 
\begin{equation}
    \frac{\partial\hat{a}^\dagger(t) }{\partial t } = \frac{1}{\sqrt{2\hbar}}\left( -\frac{\dot{\sigma}(t)}{\sigma^2(t)}\hat{q} -i\left( \dot{\sigma}(t)\hat{p}-m\ddot{\sigma}(t)\hat{q} \right)\right).
\end{equation}
From the expressions of \eqref{positionmomentumoperator} for the position and momentum operators, we find:
\begin{equation}
\begin{split}
\frac{\partial\hat{a}^\dagger(t) }{\partial t } &= \frac{1}{2} \left(-\frac{\dot{\sigma}(t)}{\sigma(t)}\left(  \hat{a}(t)+\hat{a}^\dagger(t)\right) 
- i \dot{\sigma}(t)\left( \left( M\dot{\sigma}(t)- \frac{i}{\sigma(t)}\right)\hat{a}(t) + \left( M\dot{\sigma }(t)+\frac{i}{\sigma(t)}\right)\hat{a}^\dagger(t) \right)
+ i M \sigma(t) \ddot{\sigma}(t)\left(  \hat{a}(t)+\hat{a}^\dagger(t)\right)\right)\\
&=\frac{1}{2}\left(\left(-\frac{\dot{\sigma}(t)}{\sigma(t)} +i^2\frac{\dot{\sigma}(t)}{\sigma(t)}-iM\dot{\sigma}^2(t) +iM\sigma(t)\ddot{\sigma}(t)\right)\hat{a}(t)+\left(-\frac{\dot{\sigma}(t)}{\sigma(t)} -i^2\frac{\dot{\sigma}(t)}{\sigma(t)}-iM\dot{\sigma}^2(t)+iM\sigma(t)\ddot{\sigma}(t)\right)\hat{a}^\dagger(t)\right)\\
&=\frac{1}{2}\left(\left(-2\frac{\dot{\sigma}(t)}{\sigma(t)} +iM(\sigma\ddot{\sigma}(t)-\dot{\sigma}(t)(t)^2(t))\right)\hat{a}(t)+iM(\sigma(t)\ddot{\sigma}(t)-\dot{\sigma}^2(t))\hat{a}^\dagger(t)\right).
\end{split}
\end{equation}

%%%%%%%%%%%%%%%%%%%%%%%%%%%%%%%%%%%%%%%%
\section{Derivation of \eqref{solution_phizero_integral}}
\label{app:solution_phizero}

We aim to show that
\begin{equation}\label{phaseexpressiontoevaluate}
{}_{I}\!\bra{0;t}\frac{\partial}{\partial t}\ket{0;t}_I = \int_{-\infty }^{+\infty }dq \,\phi_0^{*}(q;t)\frac{\partial \phi_0(q;t)}{\partial t}=\frac{iM}{4}\left(  \sigma(t)\ddot{\sigma}(t)-\dot{\sigma}^2(t)\right).
\end{equation}
The coordinate representation of the invariant's ground state at a  time $t$ is
\begin{equation}\label{eq:app_phi_0}
\phi_0(q;t) = \frac{1}{(\pi\hbar\sigma^2(t))^\frac{1}{4}}\exp\left( \frac{iM}{\hbar}\left(\frac{\dot{\sigma}(t)}{\sigma(t)}+\frac{i}{M\sigma^2(t)} \right)\frac{q^2}{2} \right).
\end{equation}
Taking the derivative, we find:
\begin{equation}\label{eq:app_deriv_phi_0}
\frac{\partial \phi_0(q;t)}{\partial t} = \frac{1}{(\pi\hbar\sigma^2(t))^{\frac{1}{4}}}\left( \frac{iM}{\hbar}\left( \frac{\ddot{\sigma}(t)}{\sigma(t)}-\frac{\dot{\sigma}^2(t)}{\sigma^2(t)}-\frac{2i\dot{\sigma}(t)}{M\sigma^3(t)}\right)\frac{q^2}{2}-\frac{\dot{\sigma}(t)}{2\sigma(t)} \right)\exp\left( \frac{iM}{\hbar}\left( \frac{\dot{\sigma}(t)}{\sigma(t)} +\frac{i}{M\sigma^2(t)}\right)\frac{q^2}{2}\right).
\end{equation}
If we consider (\ref{eq:app_phi_0})-(\ref{eq:app_deriv_phi_0}), \eqref{phaseexpressiontoevaluate} reads as
\begin{equation}\label{sum of two integrals}
{}_I\!\bra{0;t}\frac{\partial}{\partial t}\ket{0;t}_I = I_1+I_{2},   
\end{equation}
where
\begin{equation}
    I_1 = \frac{1}{(\pi\hbar\sigma^2(t))}\left( \frac{iM}{\hbar}\left( \frac{\ddot{\sigma}(t)}{\sigma(t)}-\frac{\dot{\sigma}^2(t)}{\sigma^2(t)}-\frac{2i\dot{\sigma}(t)}{M\sigma^3(t)}\right) \right) \int_{-\infty }^{+\infty }dq \ \frac{q^2}{2}\exp\left( -\frac{q^2}{\hbar\sigma^2(t)}\right)
\end{equation}
and
\begin{equation}
    I_2 = \frac{1}{(\pi\hbar\sigma^2(t))}\left(-\frac{1}{2}\frac{\dot{\sigma}(t)}{\sigma(t)}  \right) \int_{-\infty }^{+\infty }dq \exp\left( -\frac{q^2}{\hbar\sigma^2(t)}\right).
\end{equation}
Then, using the Gaussian integrals
\begin{equation}
    G_1(\alpha) = \int_{-\infty }^{+\infty }dy \exp\left( -\alpha y^2\right) =\sqrt{\frac{\pi
    }{\alpha}};\quad 
    G_2 = \int_{-\infty }^{+\infty }dy \ y^2 \exp\left( -\alpha y^2\right) = \frac{\sqrt{\pi}}{2}\alpha^{-\frac{3}{2}},
\end{equation}
we obtain:
\begin{equation}
    I_1 = \frac{iM}{4}\left( \ddot{\sigma}(t)\sigma(t)-\dot{\sigma}^2(t) \right)+\frac{1}{2}\frac{\dot{\sigma}(t)}{\sigma(t)} \quad \text{and} \quad I_2=-\frac{1}{2}\frac{\dot{\sigma}(t)}{\sigma(t)}.
\end{equation}
Substituting these expressions for $I_1$ and $I_2$ in \eqref{sum of two integrals}, we finally arrive at
\begin{equation}
    {}_{I}\!\bra{0;t}\frac{\partial}{\partial t}\ket{0;t}_I=\frac{iM}{4}\left( \ddot{\sigma}(t)\sigma(t)-\dot{\sigma}^2(t\right).
\end{equation}

%%%%%%%%%%%%%%%%%%%%%%%%%%%%%%%%%%%%%%%%
\section{Derivation of the phase term entering the time-dependent solution}
\label{app:alternative_derivation_phase_term}

In \ref{timedependetunitaryformalism} we have shown that through a time-dependent unitary $\hat{V}(t)$ we can make the invariant equivalent to a constant QHO with unit mass and frequency. In particular, $\hat{V}(t)=\hat{V}_2(t)\hat{V}_1(t)$ with $\hat{V}_1(t)$,$\hat{V}_2(t)$ as in \eqref{Time-dependent unitarymakingS}.
We recall that the invariant $\hat{I}(t)$ and its eigenstates $\ket{n;t}_I$ transform as
\begin{equation}\label{transformationlawsalphaequation}
    \tilde{\hat{I}} = \hat{V}(t)\hat{I}(t)\hat{V}^\dagger(t)=\frac{\hat{p}^2}{2}+\frac{\hat{q}^2}{2};
    \qquad
    \ket{\widetilde{n}}_{\tilde{I}} = \hat{V}(t) \ket{n; t}_{I}.
\end{equation}

Using \eqref{transformationlawsalphaequation}, we can rewrite the differential equations determining the phase term \eqref{Phase_ode} as
\begin{equation}\label{phaseodetransformed}
    \hbar\frac{d\alpha_{n}(t)}{dt} = {}_{\tilde{I}}\!\bra{\widetilde{n}}\left( i\hbar\hat{V}(t)\frac{\partial\hat{V}^\dagger(t)}{\partial t}-\tilde{\hat{H}}(t)\right)\ket{\widetilde{n}}_{\tilde{I}},
\end{equation}
where $\tilde{\hat{H}}(t)=\hat{V}(t)\hat{H}(t)\hat{V}^\dagger(t)$.
Thus, the determination of the phase term  has been reduced to the evaluation of the two operators on the right-hand side of the above equation.
We start by considering the first term; we take the derivative of $\hat{V}^\dagger(t)$:
\begin{equation}
    \frac{\partial\hat{V}^\dagger(t) }{\partial t}=\frac{iM}{2\hbar}\left(  \frac{\ddot{\sigma(t)}\sigma(t)-\dot{\sigma}(t)^2}{\sigma(t)^2}\right)\hat{q}^2\hat{V}^\dagger(t)-\frac{i\dot{\sigma}}{2\hbar\sigma}\hat{V}^\dagger(t)\left\{  {\hat{p},\hat{q}}\right\}.
\end{equation}
Using \eqref{eq:transformation_laws}, we finally get:
\begin{equation}
    i\hbar\hat{V}(t)\frac{\partial\hat{V}^\dagger(t)}{\partial t}=\frac{\dot{\sigma}(t)}{2\sigma(t)}\left\{  \hat{p},\hat{q}\right\}-\frac{M}{2}\left( \ddot{\sigma}(t)\sigma(t)-\dot{\sigma}^2(t) \right)\hat{q}^2
\end{equation}
The evaluation of $\tilde{\hat{H}}(t)$ is straightforward: we again use the transformation laws in \eqref{eq:transformation_laws} and obtain
\begin{equation}
    \tilde{\hat{H}}(t)=\frac{\hat{p}^2}{2M\sigma^2(t)}+\frac{M\dot{\sigma}^2(t)}{2}\hat{q}^2+\frac{\dot{\sigma}(t)}{2\sigma(t)}\left\{  \hat{p},\hat{q}\right\}+\frac{M\omega^2(t)\sigma^2(t)}{2}\hat{q}^{2}.
\end{equation}
Using the Ermakov equation \eqref{EPequation2}, we can rewrite the difference between the two operators as
\begin{equation}
     i\hbar\hat{V}(t)\frac{\partial\hat{V}^\dagger(t)}{\partial t}-\tilde{\hat{H}}(t)=-\frac{\hat{p}^2}{2M\sigma^2(t)}-\frac{M}{2}\left( \ddot{\sigma}(t)\sigma(t)+\omega^2(t)\sigma^2(t)\right)\hat{q}^2=-\frac{1}{M\sigma^2(t)}\tilde{\hat{I}}.
\end{equation}
From \eqref{transformationlawsalphaequation} we have that $\ket{\widetilde{n}}_{\tilde{I}}$ is an eigenstate of $\tilde{\hat{I}}$ with eigenvalue $\hbar(n+\frac{1}{2})$ and the evaluation of the right side of \eqref{phaseodetransformed} becomes trivial:
\begin{equation}
\frac{d\alpha_n(t)}{dt}=-\left(n+\frac{1}{2}\right) \frac{1}{M\sigma^2(t)},
\end{equation}
which coincides with \eqref{phaseodefinal} of the main text.

%%%%%%%%%%%%%%%%%%%%%%%%%%%%%%%%%%%%%%%%
\section{Analytical derivation of the transition amplitudes}
\label{APP: AnalyticMethod}

The amplitude ${}_{H}\!\bra{m;t}\ket{\psi(t)}$ can be also determined in coordinate representation, by making use of the resolution of the identity $\hat{\mathbb{I}} = \int_{}^{}dq\ket{q}\!\bra{q}$. In particular, given $\ket{\psi(t)} = \sum_{n}c_{n}\ket{\psi_n(t)}$ with $\ket{\psi_n(t)}=e^{i\alpha_{n}(t)}\ket{n;t}_I$, we compute
\begin{equation}\label{AmpitudeIntegral}
    \begin{aligned}
        {}_{H}\!\bra{m;t}\ket{\psi_n(t)} &= \int_{}^{}dq\ \Phi_m(q;t)\psi_n(q;t) \\
        &= e^{i\alpha_n(t)}\frac{1}{\sqrt{2^{m+n}m!n!}}\left( \frac{f^2(t)}{\pi^2\hbar\sigma^2(t)}\right)^\frac{1}{4} 
        \int_{}^{}dq\exp\left(-\left( \frac{M\Omega(t)}{\hbar}+f^2(t) \right)\frac{q^2}{2}\right) H_m\left(f(t)q\right)H_n\left(\frac{q}{\sqrt{\hbar}\sigma(t)}\right),
    \end{aligned}
\end{equation}
where $\Omega(t) = \frac{1}{M\sigma^2(t)}-i\frac{\dot{\sigma}(t)}{\sigma(t)}$ and $f(t)=\sqrt{\frac{M\omega(t)}{\hbar}}$.

In order to evaluate the integral
\begin{equation}\label{DoubleHermiteIntegral}
    I_{mn}(t) = \int_{}^{}dq\exp\left( -\left( \frac{M\Omega(t)}{\hbar}+f^2(t) \right)\frac{q^2}{2}\right) \ H_m\left( f(t)q \right)H_n\left( f(0)\frac{q}{\sigma(t)}\right),
\end{equation}
we use the generating function of Hermite polynomials:
\begin{equation}
    \exp\left( -s^2+2sx \right) = \sum_{k=0}^{+\infty }\frac{s^k}{k!}H_k(x).
\end{equation}
In this way, we can write the generating function of $I_{mn}(t)$, i.e.,
\begin{equation}
    \sum_{m,n=0}^{+\infty }\frac{s_1^m}{m!}\frac{s_2^n}{n!}I_{mn}(t)= \exp\Big( -(s_1^2+s_2^2)\Big) \int_{}^{}dq\exp\left( -\left( \frac{M\Omega(t)}{\hbar}+f^2(t) \right)\frac{q^2}{2} \right) \exp\left( 2\left( f(t)s_1+\frac{s_2}{\sqrt{\hbar}\sigma(t)} \right)q\right)
\end{equation}
that is a Gaussian integral resulting in
\begin{equation}\label{IntegralGeneratingFunction1}
    \sum_{m,n=0}^{+\infty}\frac{s_1^m}{m!}\frac{s_2^n}{n!}I_{mn}(t)=\sqrt{\frac{2\pi}{\frac{M\Omega(t)}{\hbar}+f^2(t)}}\exp\left( \frac{2\left( f(t)s_1+\frac{s_2}{\sqrt{\hbar}\sigma(t)}\right)^2}{\frac{M\Omega(t)}{\hbar}+f^2(t)}-(s_1^2+s_2^2)\right).
\end{equation}
Now, if we insert the explicit expression of the Bogoliubov coefficients in \eqref{BogoliubovExplicitCoefficients} and we substitute the relation $\Omega(t)=\frac{1}{M\sigma^2(t)}-i\frac{\dot{\sigma}(t)}{\sigma(t)}$, then \eqref{IntegralGeneratingFunction1} simplifies as
\begin{equation}
    \sum_{m,n=0}^{+\infty }\frac{s_1^m}{m!}\frac{s_2^n}{n!}I_{mn}(t) = \sqrt{\frac{\pi}{\frac{\sqrt{M\omega(t)}}{\hbar\sigma(t)}u(t)}}\exp\left( \frac{-v(t)s_1^2+2s_1s_2+v^*(t)s_2^2}{u}\right).
\end{equation}
Then, let us consider $m \ge n$ and use the multinomial theorem~\cite{degroot1986probability}, which we recall for clarity: for every positive integer $m$ and any non-negative integer $n$,
\begin{equation}
(x_1 + x_2 + \cdots + x_m)^n = 
\sum_{\substack{k_1 + k_2 + \cdots + k_m = n \\ k_1, k_2, \dots, k_m \geq 0}} 
\binom{n}{k_1, k_2, \dots, k_m} 
x_1^{k_1} x_2^{k_2} \cdots x_m^{k_m}
\end{equation}
where $\binom{n}{k_1, k_2, \ldots, k_m} = \frac{n!}{k_1! \, k_2! \cdots k_m!}$ is the multinomial coefficient. Hence, we get:
\begin{equation}
     \sum_{m,n=0}^{+\infty }\frac{s_1^m}{m!}\frac{s_2^n}{n!}I_{mn}(t) =\sqrt{\frac{\pi}{\frac{\sqrt{M\omega(t)}}{\hbar\sigma(t)}u(t)}} \sum_{m,n=0}^{+\infty} s_1^ms_2^n 
     \left( \frac{\left( -v(t) \right)^{\frac{m-n}{2}}}{u(t)^{\frac{m+n}{2}}}\ 2^n\sum_{k=0}^{\left[ \frac{n}{2} \right]}\frac{\left( -\left| v(t) \right|^2/4 \right)^k}{\left[ \frac{1}{2}(m-n)+k \right]! \, k! \, (n-2k)!}\right) \quad (m\ge n),
\end{equation}
from which we obtain the expression of $I_{mn}(t)$ equating the coefficients of the monomials in $s_1,s_2$. The explicit expression we obtain is:
\begin{equation}\label{eq:Imn}
    I_{mn}(t)=\sqrt{\frac{\pi}{\frac{\sqrt{M\omega(t)}}{\hbar\sigma(t)}u(t)}}  \frac{\left( -v(t) \right)^{\frac{m-n}{2}}2^nm!n!}{u(t)^{\frac{m+n}{2}}}\ \sum_{k=0}^{\left[ \frac{n}{2} \right]}\frac{\left( -\left| v(t) \right|^2/4 \right)^k}{\left[ \frac{1}{2}(m-n)+k \right]! \, k! \, (n-2k)!} \quad (m\ge n).
\end{equation}
Finally, substituting \eqref{eq:Imn}  into  \eqref{AmpitudeIntegral}, after some algebra we carry out the derivation of ${}_H\!\bra{m;t}\ket{\psi_n(t)}$ 
\begin{equation}
    \begin{aligned}
        {}_{H}\!\bra{m;t}\ket{\psi_n(t)} &= \int_{}^{}dq\ \Phi_m(q;t)\psi_n(q;t) \\
        &= e^{i\alpha_n(t)}\ u(t)^{-(n+\frac{1}{2})}\left( -\frac{v(t)}{2u(t)} \right)^{\frac{m-n}{2}}\sqrt{m!n!} \, \sum_{k=0}^{\left[ \frac{n}{2} \right]}\frac{\left( -\left| v(t) \right|^2/4 \right)^k}{\left[ \frac{1}{2}(m-n)+k \right]! \, k! \, (n-2k)!}.
    \end{aligned}
\end{equation}
The equivalence with what found through the algebraic method is achieved by setting
\begin{equation}
    \begin{cases}
        \displaystyle{ u(t)=\cosh(r(t)) \ e^{-i\chi(t)}} \\ 
        \displaystyle{ v(t)=\sinh(r(t)) \ e^{-i(\chi(t)-\phi(t))}} \\ 
        \displaystyle{ \eta(t) =\tanh(r(t)) e^{i\phi(t)}} \\ 
        \displaystyle{ \zeta(t)=2\log(\cosh(r(t)))}.
    \end{cases}
\end{equation}
The case with $m\le n$ can be done in complete analogy to the case above, starting again from \eqref{IntegralGeneratingFunction1}.

%%%%%%%%%%%%%%%%%%%%%%%%%%%%%%%%%%%%%%%%
\bibliography{bibliography}

\end{document}